\documentclass[letterpaper,twocolumn,10pt]{article}

\usepackage{usenix}
\usepackage{filecontents}

\usepackage{amsmath} % Enhances mathematical typesetting
\usepackage{amssymb} % Adds mathematical symbols
\usepackage{amsfonts} % Provides more fonts for math
\usepackage{mathtools} % Adds extensions to amsmath
\usepackage{bm} % Bold math symbols
\usepackage{bbm} % Blackboard bold for numbers and characters
\usepackage{physics} % Simplifies expressions like derivatives, integrals, etc.
\usepackage{siunitx} % SI units and scientific formatting
\usepackage{hyperref}  

\usepackage{fontawesome} % For symbols
\usepackage{xcolor}

\newcommand{\nobug}{\textcolor{green}{\faShield}} % Green shield for no violation
\newcommand{\bug}{\textcolor{red}{\faBug}} % Red bug for violation

\usepackage{amsthm} % For theorem-like environments
\newtheorem{theorem}{Theorem}
\newtheorem{lemma}{Lemma}

\newtheorem{corollary}{Corollary}
\newtheorem{definition}{Definition}

\newtheorem{assumption}{Assumption}
\usepackage[most]{tcolorbox}

\usepackage{graphicx} % For including images
\usepackage[font=normalsize]{caption} % 
\usepackage{tikz} % For creating diagrams and illustrations
\usepackage{pgfplots} % For plotting graphs
\pgfplotsset{compat=1.17}
\usepackage{subcaption}

\usepackage{booktabs} % Professional-quality tables
\usepackage{multirow} % For multi-row cells in tables
\usepackage{array} % Extends table column formatting
\usepackage{tabularx} % Tables with adjustable-width columns

\usepackage{xcolor} % Colors for text and backgrounds
\usepackage{hyperref} % Adds hyperlinks
\hypersetup{colorlinks=true, linkcolor=blue, citecolor=blue, urlcolor=blue}
\usepackage{enumitem} % Customizable lists
\setlist{nosep} % Reduce list item spacing
\usepackage{microtype} % Improves spacing and typography
\usepackage{todonotes} % For adding to-do notes
\usepackage{dirtytalk}

\usepackage{algorithm}
\usepackage[noend]{algpseudocode}

\begin{document}

\title{Monitoring Violations of Differential Privacy over Time}

\author{
{\rm Önder Askin\textsuperscript{*} } \\
Ruhr University Bochum \\
oender.askin@rub.de 
\and
{\rm Tim Kutta\textsuperscript{*}} \\
Aarhus University \\
tim.kutta@math.au.dk
\and
{\rm Holger Dette} \\
Ruhr University Bochum \\
holger.dette@rub.de
}

\maketitle

\begingroup\renewcommand\thefootnote{*}
\footnotetext{Corresponding Authors}
\endgroup

\begin{abstract}
Auditing differential privacy has emerged as an important area of research that supports the design of privacy-preserving mechanisms. Privacy audits help to obtain empirical estimates of the privacy parameter, to expose flawed implementations of algorithms and to compare practical with theoretical privacy guarantees. In this work, we investigate an unexplored facet of privacy auditing: the sustained auditing of a mechanism that can go through changes during its development or deployment. Monitoring the privacy of algorithms over time comes with specific challenges. Running state-of-the-art (static) auditors repeatedly requires excessive sampling efforts, while the reliability of such methods deteriorates over time without proper adjustments. To overcome these obstacles, we present a new monitoring procedure that extracts information from the entire deployment history of the algorithm. This allows us to reduce sampling efforts, while sustaining reliable outcomes of our auditor. We derive formal guarantees with regard to the soundness of our methods and evaluate their performance for important mechanisms from the literature. Our theoretical findings and experiments demonstrate the efficacy of our approach.

\end{abstract}

\section{Introduction}

\noindent Differential privacy (DP) \cite{Dwork_2006} is a popular approach for bounding the information leakage associated with data-processing algorithms. If used appropriately, differentially private mechanisms can ensure the privacy of individuals that are represented in a dataset. DP algorithms are therefore adopted by companies and public agencies in settings where sensitive data is involved \cite{Erlingsson_2014, Abowd_2018, Holohan_2019}. Yet, the development and deployment of DP mechanisms are susceptible to errors, which can result in algorithms that do not meet their purported level of privacy \cite{Mironov_2012,Lyu_2017,Tramer_2022}. Auditing procedures can help to detect and remedy these privacy violations, which makes them useful tools for the design and implementation of DP algorithms.

Auditing methods vary with regard to their design and objectives. Some approaches, for instance, require access to the algorithm's code \cite{StatDP,DP-Finder,Kifer_2019}, while others limit themselves to the inputs and outputs of the mechanism under investigation \cite{DP-Sniper, Dette_2022, Lu_2024}. A number of tools are designed for variants of differential privacy (rather than pure DP itself) \cite{Liu_2019, Lokna_2023, Kutta_2024}. And some are tailored to prominent mechanisms from the DP literature, such as differentially private stochastic gradient descent (DP-SGD), a cornerstone in privacy-preserving machine learning \cite{Jagielski_2020,Nasr_2023, Nasr_2025}.

A common feature of all these prior works is that auditing DP for an algorithm is treated as a one-time task. In contrast, this work considers scenarios where a data processing mechanism is regularly monitored for potential privacy violations. The need for such ongoing auditing arises naturally in the context of privacy-preserving systems.
First, the design of DP algorithms often involves extensive experimentation, making it valuable to track which modifications (inadvertently) result in privacy breaches. Second, during deployment, algorithms may undergo changes, typically aimed at improving efficiency or utility. And third, a trusted party responsible for validating the DP guarantees of these algorithms (such as a public agency or external auditor) might have an interest in continuously monitoring them over time.

In such situations, it is natural to model an algorithm as dynamically evolving over time. The task of a DP monitoring scheme would then be to continuously scan the algorithm for the emergence of potential privacy breaches. In this context, an important insight is that monitoring cannot be feasibly done with traditional auditing schemes, say by applying them over and over again as a regular screening procedure. As we discuss in this work, such an approach puts at risk the reliability of individual screenings, and more importantly, is inefficient in its data use. We therefore present a new approach that conducts regular, data-efficient "small screenings" and aggregates the ensuing information over time. While any small screening may be neither particularly reliable or sensitive, their combined (and carefully weighted) vote results in a powerful procedure that detects harmful changes shortly after they occur. The development of this procedure relies on state-of-the-art tools developed in probability theory to control the risk of false detections.

\subsection*{Main contributions} 

\noindent In this work, we present, to the best of our knowledge, the first DP auditor that is specifically designed to monitor algorithms over extended periods of time. Our method comes with \\[-1.5ex]

\begin{itemize}
        \item[(i)] significantly reduced sampling efforts, \\[-1.5ex]
        \item[(ii)] tunable confidence parameters that contain the probability of false detection over the entire time span, \\[-1.5ex]
        \item[(iii)] a mathematical theory that proves its soundness. \\[-1.5ex]
\end{itemize}

In Section \ref{Section_Background}, we discuss basics of DP and stochastic process theory. We then present our monitoring method and accompanying statistical theory in Section \ref{Section_Monitoring}. Subsequently, we evaluate our approach on common algorithms from the DP literature in Section \ref{Section_Experiments} and discuss some related works in Section \ref{Section_Related_Work}. We close with some concluding remarks in Section \ref{Section_Conclusion}.

\section{Background} \label{Section_Background}

\noindent In this section, we briefly provide background on differential privacy, sequential monitoring and stochastic processes.

\subsection{Differential Privacy}

\noindent Roughly speaking, a (randomized) algorithm $A$ is differentially private if the data of any one individual in the database $x$ does not affect its output distribution by too much. More formally, DP implies that the output distribution of $A$ on neighboring databases $x$ and $x'$ should be similar. Here, two datasets are neighbors if they differ in exactly one entry (i.e., one would obtain $x'$ from $x$ by altering the data in one row). We denote the neighborhood relation by $x  \sim_{N}  x'$.

\begin{definition} \label{Definition_DP}
An algorithm $A$ is $\varepsilon$-differentially private if for any two neighboring databases $x \sim_N x'$ and any measurable set $\mathcal{E}$ 
  \begin{align} \label{DP_equation}
\mathbb{P}\left(A(x) \in \mathcal{E}\right) \leq e^{\varepsilon} \, \mathbb{P}\left( A(x') \in \mathcal{E}\right).
\end{align}
\end{definition}

If the privacy parameter $\varepsilon > 0$ is chosen small enough, an adversary observing $A$'s output will not be able to confidently decide if database $x$ or $x'$ was used to produce said output. Therefore, assuming that each row of $x$ corresponds to the data provided by one individual, an adversary cannot reliably infer membership in the dataset from the output.

To distinguish DP from its variants, Definition \ref{Definition_DP} is usually called \textit{pure differential privacy}. The definition also serves as a useful basis for auditing the privacy of an algorithm under investigation. More concretely, prior works have relied on this definition to test the privacy claims of a given mechanism \cite{StatDP,DP-Finder, CheckDP}. Vulnerable algorithms are found with the help of a suitable tuple $(x_0,x'_0,\mathcal{E}_0)$ that provokes a privacy violation of the form
\begin{align} \label{DP_violation_Eq}
\mathbb{P}\left(A(x_0) \in \mathcal{E}_0\right) > e^{\varepsilon} \, \mathbb{P}\left( A(x'_0) \in \mathcal{E}_0\right).
\end{align}
If, on the other hand, no violation as in \eqref{DP_violation_Eq} is observed for the tuple, then no breach of privacy is reported. A reliable auditing procedure will therefore also depend on a suitable choice of these tuples. While simple heuristics often suffice to find these \cite{StatDP}, there also exist tools dedicated to their search  \cite{DP-Sniper}. We discuss suitable tuples for specific algorithms in our experiments in Section \ref{Section_Experiments}. 

In our work, we also adopt an approach that leans on Definition \ref{Definition_DP} to check for privacy violations (see Section \ref{Section_Monitoring}), with the difference that our auditor and privacy condition are formulated for sequential monitoring over time.

\subsection{Sequential monitoring} \label{sec:far}

\noindent Sequential monitoring refers to the setting where an analyst observes data points $R_1, R_2, R_3, \ldots$ sequentially and looks for the emergence of a structural change over time. At any time $\tau$, the analyst combines the available data ${R_1, \ldots, R_\tau}$ into a detector $\widehat{D}(\tau)$. When the next observation $R_{\tau+1}$ arrives, the detector is updated to $\widehat{D}(\tau+1)$, and so forth. A large detector value indicates that a change has taken place. More precisely, the analyst compares $\widehat{D}(\tau)$ to a threshold $q$ and if
\begin{align} \label{e:det:con}
\widehat{D}(\tau) > q,
\end{align}
the analyst decides that a change has occurred and stops. Otherwise, if \eqref{e:det:con} is not satisfied, no change is detected and monitoring continues until a terminal time $T$. A reliable monitoring method controls the false alarm rate (FAR), that is, the probability of incorrectly declaring a change when none has occurred. The FAR can be controlled by adjusting the threshold $q$, with larger values reducing the FAR but also making it harder to detect actual changes. It is common in the literature to set the FAR to an acceptable error level $\alpha \in (0,1)$, typically $0.05$. Since exactly obtaining FAR = $\alpha$ is difficult, most procedures only approximate this level for large $T$. For a detailed review on sequential monitoring, we refer the reader to \cite{aue:kirch:2024}. The study of sequential monitoring (and more precisely the choice of an adequate threshold $q$) is closely connected to the theory of stochastic processes and most importantly to characteristics of the Brownian motion, discussed next.

\subsection{The standard Brownian motion} \label{sec:brown}

\noindent A stochastic process $S$ indexed in an interval $I$ is a random function $S: I \to \mathbb{R}$. This means that for any fixed value $u \in I$ the value $S(u)$ is a random variable. If $S$ is continuous (for any random outcome), $S$ can be characterized using its marginal distributions, i.e., distributions of the vector $(S(u_1),\cdots,S(u_R))$ for collections of points $u_1,\cdots,u_R \in I$. If these vectors follow Gaussian distributions, then $S$ is called a Gaussian process.

\begin{figure}[ht]
\centering
\includegraphics[width=0.4\textwidth]{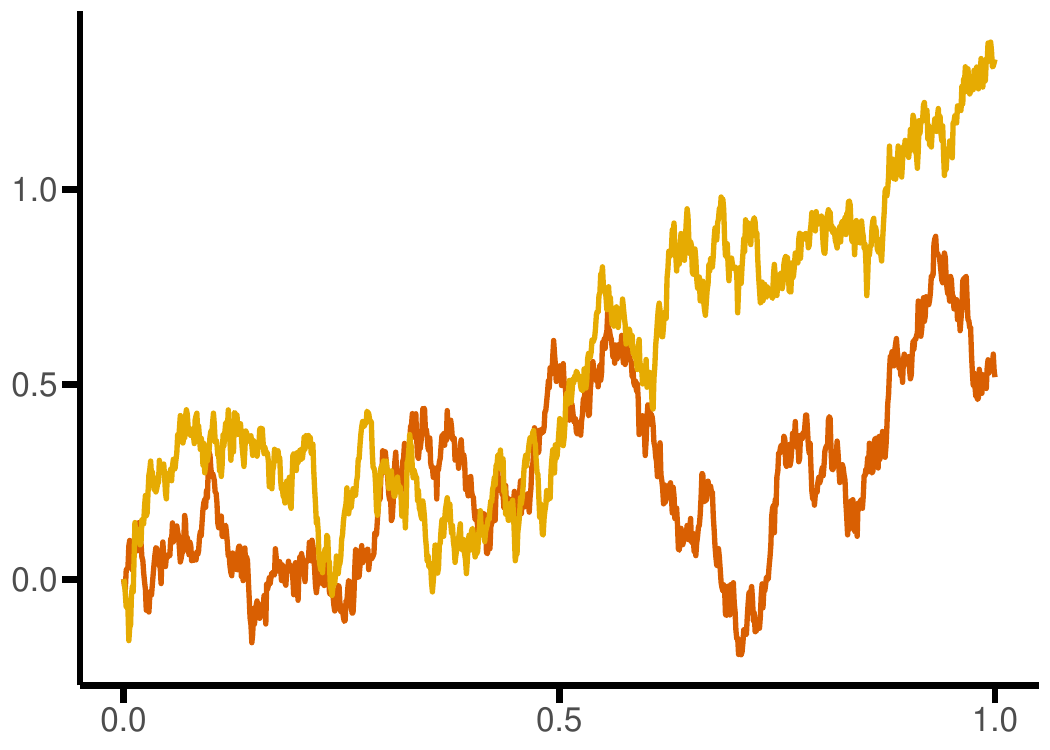}
\caption{Two realizations of the standard Brownian motion.}
\label{fig:brownian}
\end{figure}

The single most important Gaussian process in applied mathematics is the standard Brownian motion (sBm), which is indexed in $I=[0,1]$. 
At each point $u$, $B(u)$ follows a normal distribution $\mathcal{N}(0, u)$, meaning it has mean zero and variance $u$. The behavior of the sBm is fully characterized by its covariance structure for $u,v \in [0,1]$:
\[
\mathbb{C}ov(B(u), B(v)) = \min(u,v).
\]
 For a formal introduction to the sBm and its key properties, we refer to \cite{karatzas:shreve:1991}. 
The sBm was first used in physics to describe the random motion of particles in a fluid, a phenomenon now called "Brownian motion". Since then, it has become fundamental in areas such as economics, biology, and quality control.
Mathematically, the sBm can be viewed as a standard Gaussian distribution that outputs random functions, rather than just random numbers. Just as the central limit theorem describes the convergence of sums of random numbers to a Gaussian distribution, there is a version for random functions where the limit is the sBm (see, e.g., \cite{vaart:wellner:1996}).
To develop some intuition, we show two realizations of the sBm (generated with different random seeds) in Figure \ref{fig:brownian}.

\section{Privacy monitoring for an evolving algorithm} \label{Section_Monitoring}

\noindent \textbf{Overview} In this section, we present a model for a randomized algorithm $A$ that evolves over time. We treat time as a discrete variable, spanning $t = 1, 2, \ldots, T$, where each time point $t$ may represent an hour, day, or week. Our focus is on real-time auditing of the differential privacy of $A$. Specifically, we consider the problem of continuously monitoring the outputs of $A$ to detect privacy violations that may result from potential implementation bugs.
Existing approaches to DP auditing are designed for fixed algorithms. A straightforward approach to monitoring $A$ might involve repeatedly applying a standard DP-auditor at each time $t$ across the entire time span. However, as we will demonstrate below, this naive approach is inefficient and leads to significant oversampling. 
To address this, we develop a new time-aggregator that carefully combines a sequence of naive auditors over time into a more powerful procedure. The methodology is detailed in Section \ref{s:meth}. In Section~\ref{s:theo}, we present mathematical theory, providing performance guarantees for our approach. Our construction uses state-of-the-art tools from stochastic process theory that help to reduce sampling effort, while obtaining reliable decisions with regards to privacy.

\subsection{Outline of the auditing methodology}\label{s:meth}

\noindent \textbf{Model for an evolving algorithm}
For our theoretical analysis, we consider an algorithm $A$ that evolves over discrete time $t = 1, \ldots, T$. The output distributions of $A$ may change with time, and we explicitly denote this dependence by writing $A_t$ (the algorithm at time $t$). In our model, the algorithm may undergo modifications by the owner, and we denote the time points of these modifications by $t_k$, where $1 < t_1 < t_2 < \cdots < t_K < T$. Here, $K \in \mathbb{N}_0$ represents the total number of modifications, which may be zero.
The relationship between these modifications can be expressed as follows:
\begin{align} \label{e:alg:mod}
& A_1 = A_2 = \cdots = A_{t_1 - 1} \\
\neq & A_{t_1} = A_{t_1 + 1} = \cdots = A_{t_2 - 1} \nonumber\\
\neq & \quad \cdots \nonumber\\
\neq & A_{t_K} = A_{t_K + 1} = \cdots = A_T. \nonumber
\end{align}
In our theory, the analyst does not necessarily have to know the times or number of these modifications. \\
%In the setting of this paper, the analyst does not know the times or number of these modifications.

\noindent \textbf{The NPVT condition} Recall Definition \ref{Definition_DP} for the rigorous formulation of DP for a fixed algorithm. 
We now specify what it means for $(A_t)_{t=1,\cdots,T}$ to satisfy differential privacy (DP) over the entire time period. Consider a pair of neighboring databases $x \sim_N x'$ and a potentially vulnerable event $\mathcal{E}$, which represents a specific privacy risk. If $A_t$ satisfies $\varepsilon$-DP at all times $t$, it must hold that
\begin{align} \label{e:DP:cond}
p_t := \mathbb{P}(A_t(x) \in \mathcal{E}) - e^\varepsilon\,\, \mathbb{P}(A_t(x') \in \mathcal{E}) \leq 0 \quad \forall t.
\end{align}
We can formulate this more compactly in the following "\textit{no privacy violation over time}" (NPVT) condition
\begin{align} \label{e:NPVT}
\textnormal{NPVT:} \qquad\quad\max_{1 \leq t \leq T} p_t \leq 0.
\end{align}

\noindent \textbf{The naive auditor}  
We now address the task of monitoring $A$ over time to detect implementation errors as they arise. Specifically, the analyst seeks to identify breaches of the NPVT condition in \eqref{e:NPVT}. At any time $t$, the analyst can execute the algorithm $A_t$ with input databases $x$ and $x'$ to produce independent outputs. Denote these outputs as $X_{1,t}, \ldots, X_{n,t} \sim A_t(x)$ and $Y_{1,t}, \ldots, Y_{n,t} \sim A_t(x')$, where $n$ is the number of runs.  
Using the datasets $\{X_{1,t}, \ldots, X_{n,t}\}$ and $\{Y_{1,t}, \ldots, Y_{n,t}\}$, the analyst can compute an empirical approximation of $p_t$ as follows
\begin{align}
    n_t^X := &|\{X_{i,t} \in \mathcal{E}: 1 \le i \le n\}|,\nonumber\\   
n_t^Y := &|\{Y_{i,t} \in \mathcal{E}: 1 \le i \le n\}|, \quad \nonumber\\
\hat{p}_t := &\frac{n_t^X - e^\varepsilon n_t^Y}{n}.
\end{align}
This estimator $\hat{p}_t$ has been used, for instance, in the seminal work of \cite{StatDP} to approximate $p_t$. Moreover, it allows the construction of a confidence interval for $p_t$, forming the basis for an auditing procedure (all technical claims on the naive procedure are discussed in Appendix \ref{sec:naive}).  
An accurate estimator of the variance of $\hat{p}_t$ is given by 
\begin{align} \label{e:hat:sig}
    \hat{\sigma}_t^2 := \frac{n_t^X}{n^2} \left(1 - \frac{n_t^X}{n}\right) + e^{2\varepsilon} \frac{n_t^Y}{n^2} \left(1 - \frac{n_t^Y}{n}\right).
\end{align}
Using the normal approximation of binomial random variables, we have
\begin{align} \label{e:pt/s}
\frac{\hat{p}_t - p_t}{\hat{\sigma}_t} \approx \mathcal{N}(0,1),
\end{align}
where the right-hand side denotes a standard normal distribution. For a level $\alpha \in (0,1)$, a one-sided confidence interval for $p_t$ is
\begin{align} \label{e:Ct}
\mathcal{C}_t(\alpha) := \big(\hat{p}_t+\Phi^{-1}(\alpha)\hat{\sigma}_t , \infty\big), 
\end{align}
where $\Phi^{-1}$ is the quantile function of the standard normal. A privacy violation is detected if $0 \notin \mathcal{C}_t(\alpha)$. The logic of this decision is based on the fact that $p_t \in \mathcal{C}_t(\alpha)$ with high probability. If $0 \notin \mathcal{C}_t(\alpha)$, it must be because $0 < p_t$, indicating a breach of the DP condition \eqref{e:DP:cond}. The auditor operates with an error level $\alpha$, meaning it will only erroneously detect a violation with probability $\alpha$ if DP holds.  \\
As a straightforward way to extend this method to NPVT auditing (see \eqref{e:NPVT}), one might examine whether  
\begin{align} \label{e:naive}
0 \in \bigcap_{t=1}^T \mathcal{C}_t(\alpha/T).  
\end{align} 
Here, the confidence intervals $\mathcal{C}_t(\alpha/T)$ are adjusted using the Bonferroni correction to maintain an overall level of $\alpha$ (this is the FAR from Section \ref{sec:far}).  While intuitive, this approach is inefficient for two key reasons. First, the Bonferroni correction significantly widens the confidence intervals $\mathcal{C}_t(\alpha/T)$, reducing the auditor’s ability to detect DP violations, especially for large $T$. Second, the naive auditor uses only the samples $X_{1,t}, \ldots, X_{n,t}$ and $Y_{1,t}, \ldots, Y_{n,t}$ collected at time $t$, ignoring historical data from $t=1, 2, \ldots, t-1$. This results in a substantial loss of information.
Taking these two problems together, the naive auditor can only work for very large sizes of $n$, making such a procedure computationally expensive (and for complex algorithms probably infeasible). For this reason we propose a different procedure that carefully aggregates data over time to extract a maximum amount of information. \\

\noindent \textbf{The weighted aggregator} The naive auditor presented in the previous paragraph is based on the statistic $\hat p_t/\hat \sigma_t$. To see this, notice that no DP violation is detected if
\begin{align} \label{e:naive:des}
0 \in \mathcal{C}_t(\alpha/T) \,\,\,\Leftrightarrow\,\,\, \frac{\hat p_t}{\hat \sigma_t} <-\Phi^{-1}(\alpha/T).  
\end{align}
Conversely, a large enough value of $\hat p_t/\hat \sigma_t$ implies that $0 \not\in \mathcal{C}_t(\alpha/T)$, indicating a DP violation. Our main idea is to combine these statistics over time to improve joint performance. The mathematical underpinnings of this analysis stem from the field of stochastic process theory, which lies at the heart of many modern statistical applications.  \\
Recall that, at any time $\tau$, the analyst has access to the historical sequence $\hat p_1/\hat \sigma_1, \hat p_2/\hat \sigma_2, \ldots, \hat p_\tau/\hat \sigma_\tau$. We propose the following aggregation scheme, parameterized by $\beta \in [0, 1/2)$ (for a recommended choice, see the end of this section):
\begin{align} \label{e:DT}
    \widehat{D}(\tau) := \max_{0 \leq \ell < \tau} \bigg[\frac{1}{(\ell+1)^\beta \cdot T^{1/2-\beta}} \sum_{t=\tau-\ell}^\tau \frac{\hat p_t}{\hat \sigma_t}\bigg].
\end{align}
A privacy violation is detected if $\widehat{D}(\tau)$ exceeds a critical threshold (details in the next paragraph).
To understand $\widehat{D}(\tau)$, consider specific values of $\ell$ in \eqref{e:DT}. For $\ell = 0$, the aggregation reduces to
\begin{align} \label{e:ell=0}
    \frac{1}{(0+1)^\beta \cdot T^{1/2-\beta}} \sum_{t=\tau}^\tau \frac{\hat p_t}{\hat \sigma_t} = \frac{1}{T^{1/2-\beta}} \frac{\hat p_\tau}{\hat \sigma_\tau}.
\end{align}
Ignoring the scaling factor $T^{1/2-\beta}$, this is equivalent to the naive privacy estimator at time $\tau$. The strength of this estimator is that it only relies on the most recent data. Its weakness is that it ignores potentially useful information from earlier times ($t < \tau$). If the right side of \eqref{e:ell=0} is sufficiently large, a privacy violation is detected, and the procedure stops. Otherwise, the aggregator starts incorporating historical data by increasing $\ell$. For $\ell = 1$, the aggregated statistic becomes
\[
\frac{1}{(1+1)^\beta \cdot T^{1/2-\beta}} \sum_{t=\tau-1}^\tau \frac{\hat p_t}{\hat \sigma_t} = \frac{1}{2^\beta \cdot T^{1/2-\beta}} \bigg[\frac{\hat p_{\tau-1}}{\hat \sigma_{\tau-1}} + \frac{\hat p_\tau}{\hat \sigma_\tau}\bigg].
\]
Thus, historical data from time $\tau-1$ is added to see if this strengthens the evidence to conclude that a privacy violation has occurred. The data from time  $\tau-1$ potentially helps the auditor by reducing its variance. However, it is also not the latest data anymore - if a violation has only just occurred at time $\tau$ looking at the (private) history will not help at all. The ideal cut-off $\ell$ includes all recent helpful data while ignoring all prior, now obsolete, historical information.
Since this "sweet spot" for $\ell$ is unknown, the procedure evaluates all possible values, combining them in $\widehat{D}(\tau)$. The final decision rule identifies $\widehat{D}(\tau)$ as unexpectedly large when it exceeds a predetermined threshold (indicating a privacy violation). The method to set this threshold is specified in the next paragraph. \\

\noindent \textbf{Auditing via thresholding} To summarize our auditing procedure briefly: At each time point $\tau$, the user computes $\widehat{D}(\tau)$ and checks whether it exceeds a threshold $q$. If it does, a privacy violation is detected; otherwise, the auditing continues without detecting a violation. A formal description of this process is given in Algorithm \ref{alg:dp_auditing}. The choice of the threshold $q = q(\alpha)$ determines the FAR of the procedure (see Section \ref{sec:far}), i.e., the risk $\alpha$ of falsely detecting a violation when none exists. If the naive procedure in \eqref{e:naive} were used, the threshold would be 
$q(\alpha) = -\Phi^{-1}(\alpha/T)$ (see \eqref{e:naive:des}). This is a quantile of the standard normal distribution. Similarly, for our aggregated statistic, $q(\alpha)$ is defined as the (upper) $\alpha$-quantile of a more complex probability distribution involving the standard Brownian motion (sBm), a continuous-time stochastic process $B:[0,1] \to \mathbb{R}$. An introduction to the sBm can be found in Section \ref{sec:brown}. Using the sBm, we define the random variable
\begin{align} \label{e:def:D}
&D := \sup_{0 \le u < v \le 1} d(u,v), \\
& \,\,\,\text{where} \;\,\, d(u,v) := \frac{B(u) - B(v)}{(u-v)^{\beta}}, \nonumber
\end{align}
with parameter $\beta \in [0,1/2)$ (the same parameter as in the definition of $\widehat{D}$ in \eqref{e:DT}). Setting $q(\alpha)$ as the upper $\alpha$-quantile of $D$ provides an appropriate threshold to guarantee the confidence level $\alpha$. The connection between $\widehat{D}$ and $D$ is discussed in greater detail in the Appendix (see Lemma \ref{lem:_help}). Roughly speaking, $\widehat{D}$ can be expressed as maximizing the weighted increments of a stochastic process and it converges to the maximum of weighted increments of the sBm, i.e. to $D$.
In practice, computing $q(\alpha)$ is feasible via Monte Carlo simulation. Importantly, this threshold can be determined prior to auditing since it does not depend on the specific output distributions of $A_t$.

\begin{figure}[ht]
    \centering
    \includegraphics[width=0.5\textwidth]{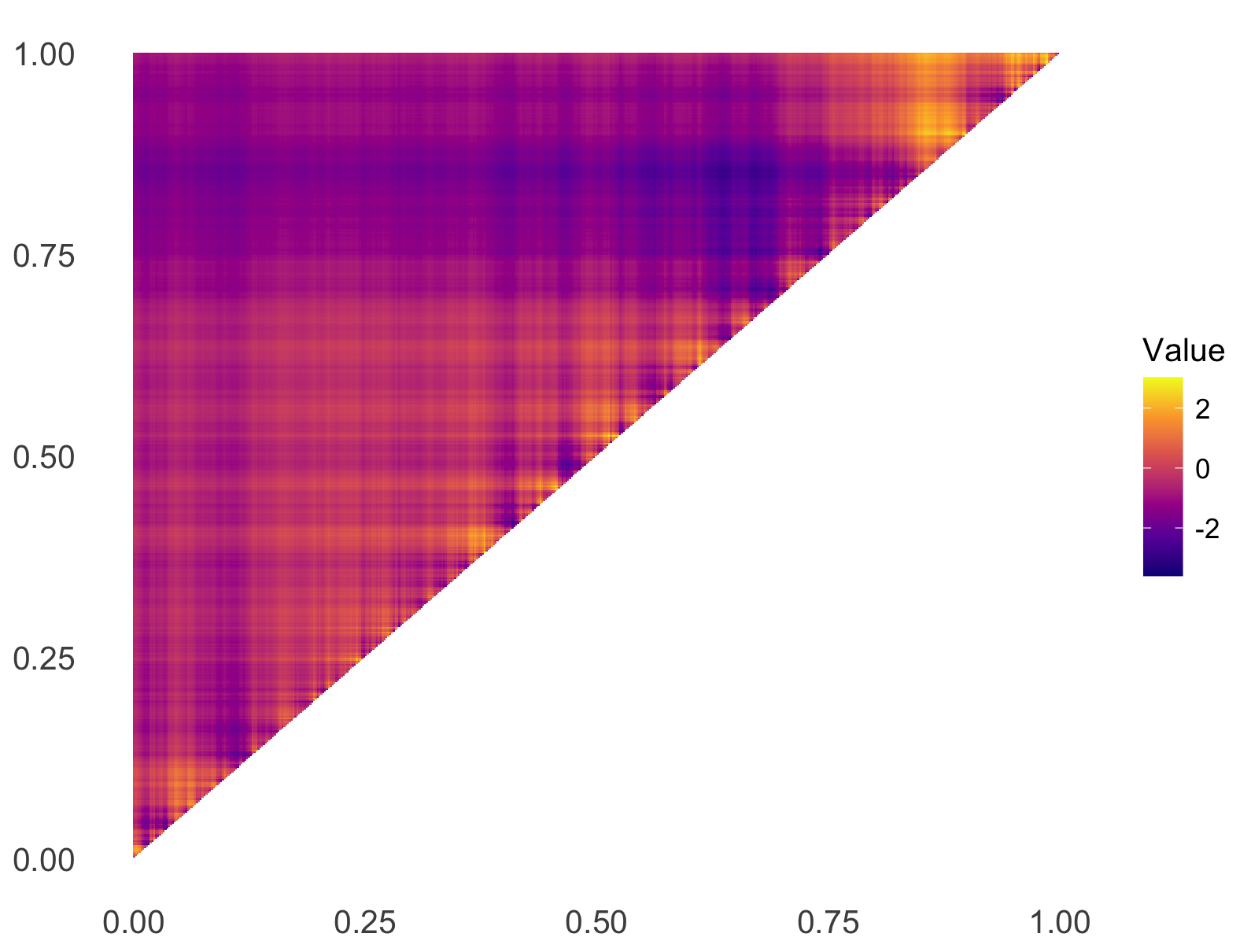}
    \caption{A sample realization of the function $d(u,v)$ defined in \eqref{e:def:D} over the domain $0 \le u < v \le 1$. This illustrates the stochastic behavior of the scaled Brownian increments used in threshold calibration.}
    \label{fig_d(u,v)}
\end{figure}

\begin{algorithm}
\caption{ReTiDP: Real-Time DP Auditing}\label{alg:dp_auditing}
\begin{algorithmic}[1]
\Require $\varepsilon$ (targeted DP level), $\alpha$ (nominal level), $\beta$ (weight parameter), $n$ (sample size), $T$ (time horizon), $\tau$ (current time), access to the inputs and outputs of $(A_t)_{1 \leq t \leq \tau}$
\Ensure Auditing decision on DP violation

\State Compute $\hat{p}_t / \hat{\sigma}_t$ for all $1 \leq t \leq \tau$
\State Compute $\widehat{D}(\tau)$ using the formula defined in eq.~\eqref{e:DT}
\State Compute the upper $\alpha$ quantile $q(\alpha)$ of $D$
\If{$\widehat{D}(\tau) > q(\alpha)$}
    \State \textbf{Return} "DP violation detected"
\Else
    \State \textbf{Return} "No DP violation detected"
\EndIf
\end{algorithmic}
\end{algorithm}

Our auditor in Algorithm \ref{alg:dp_auditing} can be used as an extension of static tools, especially ones that also check for violations of DP with tuples $(x,x',\mathcal{E})$. And since our method, in principle, only requires access to the inputs and outputs of $A$, it is compatible with approaches that work in the white- \cite{DP-Finder}, semi-black- \cite{StatDP} and black-box setting \cite{DP-Sniper}. We conclude this section by giving additional details about our procedure. \\

\noindent \textbf{Variance stabilization} For certain (extreme) choices of $\mathcal{E}$ it may happen that the variance estimator $\hat \sigma_t$ in \eqref{e:hat:sig} equals $0$. In this case, by construction, also $\hat p_t=0$ and we may interpret the ratio $\hat p_t/\hat \sigma_t=0/0=0$. However, even non-zero but very small values of $\hat \sigma_t$ can cause numerical instabilities. For this reason it can make sense to use the stabilized variance estimate
    \begin{align} \label{e:stab}
        \tilde \sigma_t(c):= \max(\hat \sigma_t, c)
    \end{align}
    for some small, positive constant $c$. It turns out that our theoretical guarantees remain valid for suitable choices of $c$ (see Corollary \ref{e:cor:stab} below).\\
    
\noindent \textbf{Weight parameter $\beta$} Our procedure involves a user-determined weight parameter $\beta \in [0,1/2)$ that occurs in the definition of $\widehat{D}$ and $D$. Roughly speaking, $\beta$ implies a trade-off between finding larger privacy violations faster ($\beta$ closer to $1/2$) and finding smaller violations more reliably ($\beta$ closer  to $0$). A similar trade-off has been studied in statistical monitoring; see e.g. \cite{kutta:doernemann:2025}. Our simulations suggest that choosing $\beta=1/4$ right in the middle of the permissible range produces good overall results and we recommend this choice to users.

\subsection{Theoretical analysis}  \label{s:theo}

\noindent \textbf{Model specifications} In this section we give mathematical performance guarantees for our monitoring scheme. For this purpose, recall
the model \eqref{e:alg:mod} and the (to the analyst unknown) modification times $1 < t_1<...<t_K< T$. To formulate mathematical theory it is easiest to express the times of changes in terms of proportions of $T$ and we hence set for some proportions $0<\theta_1<...<\theta_K<1$ 
\[
t_i = \lfloor T \theta_i \rfloor, \quad i=1,...,K.
\]
Here, the expression $\lfloor x \rfloor$ refers to the largest integer that is smaller than a number $x$ (floor function). We now state the main theoretical assumption of our analysis. 

\begin{assumption}  \textnormal{\label{ass:1}
In our asymptotics, we let $n \to \infty$. The number of time points $T=T(n)$ may be increasing in $n$. Specifically, there exists a constant $\rho \ge 0$, such that $T(n)/n^\rho$ converges to a positive real number as $n \to \infty$.}
\end{assumption}
Assumption \ref{ass:1} is extraordinarily mild. It essentially states that $T$ can be much smaller than $n$, of the same size or much larger. It only cannot be exponentially larger than $n$. This restriction is practically always fulfilled because DP auditing typically requires $n$ to be larger than $T$ anyways (this will be true in all of our experiments). Also notice that the case where $T$ is fixed and only $n \to \infty$ is covered by our theory. \\

\noindent \textbf{Main results} The following main mathematical result has two parts. First, it shows that if the NPVT condition holds (see eq. \eqref{e:NPVT}), then the probability of false detection is bounded for large samples by $\alpha$. The second part states that if NPVT is violated and there is a privacy violation, the probability of detecting this privacy violation goes to $1$. In the results we let $n \to \infty$ but remember that $T=T(n)$ may be increasing as specified by Assumption \ref{ass:1}.
\begin{theorem} \label{theo:main}\textnormal{
    Suppose that Assumption \ref{ass:1} holds and denote by $O$
 the output of Algorithm \ref{alg:dp_auditing} for some level $\alpha \in (0,1)$.
 \begin{itemize}
     \item[i)] If NPVT in \eqref{e:NPVT} holds, it follows that
     \begin{align*}
         \limsup_{n \to \infty} \mathbb{P}\big(O = \textnormal{"DP violation detected"}\big) \le \alpha.
     \end{align*}
    \item[ii)] If NPVT in \eqref{e:NPVT} does not hold, it follows that
     \begin{align*}
         \lim_{n \to \infty} \mathbb{P}\big(O = \textnormal{"DP violation detected"}\big) =1.
     \end{align*}
 \end{itemize}
 }
 \end{theorem}

 We also provide a small corollary on the variance stabilization.

 \begin{corollary}\label{e:cor:stab}\textnormal{
     Suppose that Assumption \ref{ass:1} holds and let $(c_n)_n$ be a sequence of positive numbers such that $c_n \downarrow 0$. Then, the statements of Theorem \ref{theo:main} remain true, if one replaces $\hat \sigma_n$ in the definition of $\widehat{D}$ by $\tilde \sigma_n(c_n)$ (defined in eq. \eqref{e:stab}).}
 \end{corollary}

\noindent \textbf{Detection delay} Part ii) of Theorem \ref{theo:main} implies that (asymptotically) any violation of the NPVT condition will be detected by our procedure. A lingering question is how long it takes after a "bad modification" to detect it. More precisely, let us suppose that NPVT is violated and the violation occurs first after the modification $t_{k^*}$ for some $k^* \in \{1,...,K\}$. We denote by $\tau^*$ the first time $\ge k^*$, where Algorithm \ref{alg:dp_auditing} gives out "DP violation detected". A relevant question for the user is how large the difference $\tau^*-t_{k^*}$ is, i.e. how long after a bug is introduced it takes to detect it. The gap between signal and detection is canonically called "detection delay" in the literature \cite{aue:kirch:2024} and 
shorter delays are preferable. In the case of DP monitoring a shorter delay means that an algorithmic bug is faster exposed and thus the deployment of a faulty algorithm can be stopped faster. For the statement of the next result, we need the stochastic Landau symbol denoted by  "$\mathcal{O}_P$". Basically for a sequence of positive numbers $(a_n)_n$ the symbol $\mathcal{O}_P(a_n)$ refers to a product of $a_n \cdot X$ where $X$ is a random variable (i.e. we have a rate of $a_n$ times a random factor that does not depend on $n$).  

\begin{theorem} \label{e:theo:delay}
    Suppose that Assumption \ref{ass:1} holds and that the NPVT condition in \eqref{e:NPVT} is violated. Then, it follows for the detection delay that
    \[
    \tau^*-t_{k^*} = \mathcal{O}_P\bigg(\frac{T^{(1/2-\beta)/(1-\beta)}}{n^{1/(2(1-\beta))}}\bigg).
    \]
\end{theorem}

Theorem \ref{e:theo:delay} states that delay times are very short (in particular much shorter than any fixed fraction of $T$). Larger values of $\beta$ imply shorter detection delays, at the cost of slightly worse approximations for the level $\alpha$. Larger values of $n$ imply shorter detection delay, which makes sense because they mean larger samples and hence more precise estimates.

\section{Experiments} \label{Section_Experiments}
\noindent We evaluate our monitoring framework by applying it to algorithms from the DP literature that are known to be susceptible to privacy bugs. Our experiments show that our methods can quickly detect flawed variants and changes that result in a privacy breach.

\subsection{Algorithms}
\noindent \textbf{Laplace Mechanism} Additive noise algorithms like the Laplace mechanism can be used to privatize statistics and information extracted from a database. For instance, we can consider the sum statistic 
\begin{align*}
    S(x) = \sum\limits_{i=1}^m x_i
\end{align*}
for a database $x = (x_1, ..., x_m)$ with $x_i \in [0,1]$ for each $i \in \{ 1, ...,m \}$. The output $S(x)$ can then be privatized by adding Laplace noise calibrated to its sensitivity
\begin{align*}
    \triangle_S := \max_{x \, \sim_{N} \, x'} \left\vert S(x) - S(x') \right\vert.
\end{align*}
The sensitivity of $S$ on the unit cube $[0,1]^m$ is 1  and the Laplace mechanism, defined as 
\begin{align} \label{Laplace_mechanism}
    A(x) = S(x) + Z
\end{align}
for $Z \sim Lap(0,b) $, achieves a privacy level of $\varepsilon = \frac{1}{b}$. For our correct implementation of the Laplace mechanism, we choose $b = 1$, which yields $\varepsilon = 1$. We then consider two changes to $A$ that adversely affect its privacy. The first is to choose a smaller noise scale $b = 0.5$, which only leads to a privacy level $\varepsilon' = 2$ instead of the targeted $\varepsilon = 1$. The second is to replace the Laplace noise in \eqref{Laplace_mechanism} entirely with noise from a Gaussian distribution that has the same mean and variance as the original perturbation term. In our setting, this means adding $Z \sim \mathcal{N}(0, 2 b^2)$ to $S(x)$ with $b = 1$. Algorithm $A$ in \eqref{Laplace_mechanism} would then not satisfy pure DP as in Definition \ref{Definition_DP}, but merely approximate DP with parameters $\varepsilon = 1$ and $\delta \approx 0.0396$. To monitor the privacy of the Laplace mechanism, we choose neighboring databases
\begin{align*}
    x = (0,0,...,0) \quad \text{and} \quad x' = (1,0,...,0)
\end{align*}
with $m = 10$ and events of the form $\mathcal{E} = (-\infty,a]$ (we choose $a = 0$ for the change in noise scale and $a = -1$ for the change in noise type). \\

\noindent \textbf{Report Noisy Max} Many DP algorithms operate on queries $Q_i$ evaluated on the database $x$, rather than the data itself. The Report Noisy Max algorithm (RNM), for instance, adds Laplace noise $Z_i \sim Lap\big(0,\frac{2 \, \triangle_{Q_i}}{\varepsilon}\big)$ to each query in
\begin{align*}
    Q(x) = \left( Q_1(x), ..., Q_d(x) \right)
\end{align*}
and returns the index of the largest noisy query answer, i.e., 
\begin{align*}
    \textnormal{arg max}_{i \in \{1, \cdots, d\} }(Q_1(x) + Z_1, ..., Q_d(x) + Z_d).
\end{align*}
RNM achieves a privacy level of $\varepsilon$ in this configuration. Note that there is a \say{information minimization} principle involved \cite{Dwork_2014}. A single change in the database $x$ could, in theory, affect each query $Q_i$ and therefore inflate the associated privacy loss. However, since only the index of the largest noisy query is reported (as opposed to the entire noisy vector), the privacy of RNM remains stable. Accordingly, a flawed version of RNM studied in the literature \cite{StatDP} falsely reports the largest noisy value
\begin{align*}
    \textnormal{max}_{i \in \{1, \cdots, d\} }(Q_1(x)+Z_1, ..., Q_d(x)+Z_d)
\end{align*}
and thus breaks the promised privacy level $\varepsilon$. In our experiment, we study the change from the correct version of RNM (with parameter $\varepsilon = 1$) to this incorrect version. Similar to prior work, we focus on counting queries with sensitivity $\triangle_{Q_i} = 1$ and choose
\begin{align*}
    Q(x) = (1,1,1,1,1) \quad \text{and} \quad Q(x') = (2,2,2,2,2),
\end{align*}
as well as $\mathcal{E} = (-\infty, a]$ with $a = 2$ for our auditing procedure. \\

\noindent \textbf{Sparse Vector Technique} Another prominent algorithm that operates with noisy queries is the Sparse Vector Technique (SVT). Here, queries are again processed with Laplace noise and then checked against a noisy threshold. It is then reported if a noisy query lies above (output 1) or below (output 0) this threshold. The number of query answers above the threshold is limited by a bound $c \in \mathbb{N}$ and SVT stops processing queries once this limit is reached. In \cite{Lyu_2017}, several correct and erroneous versions of SVT are examined. We adopt the numbering from \cite{Lyu_2017} while referring to these variants of SVT and choose SVT 2, a correct version that achieves its targeted privacy level $\varepsilon$, as a starting point. We assume that each query has the same sensitivity $\triangle$ and point to Algorithm \ref{Alg_SVT_2} for a more detailed outline of SVT 2.

\begin{algorithm}[ht]
    \caption{SVT 2 (from \cite{Lyu_2017})}
    \label{Alg_SVT_2}
    
    \begin{algorithmic}[1]
        \Require $x$ (database), $Q_1, ..., Q_d$ (queries), $\Gamma$ (threshold), \mbox{\qquad \,} $\varepsilon$ (target privacy level),  $c \in \mathbb{N}$ (bound)
        \Ensure $y_1, y_2, \cdots$ (binary sequence)
        \State Set count $ \gets 0 $
        \State Sample $Z \sim Lap(0, 2c \; \triangle / \varepsilon)$
        \For{$i = 1, ..., d$}
        \State Sample $Z_i$ with $Z_i \sim Lap(0, 4c \; \triangle / \varepsilon)$
        \If{$Q_i(x) + Z_i \geq \Gamma + Z$}
            \State \textbf{Output} $y_i = 1$
            \State Sample $Z \sim Lap(0, 2c \; \triangle / \varepsilon) $
            \State Set count $\gets$ count + 1
            \If{count $\geq c$}
                \State \textbf{Abort}
            \EndIf
        \Else 
            \State \textbf{Output} $y_i = 0$
        \EndIf
        \EndFor
    \end{algorithmic}
\end{algorithm}

\begin{figure*}
\begin{subfigure}[c]{.49\linewidth}
\centering
\caption{\textbf{Laplace Mechanism - Change in Noise Scale \,\,\bug}}
\includegraphics[width=0.95\linewidth,height=140pt]{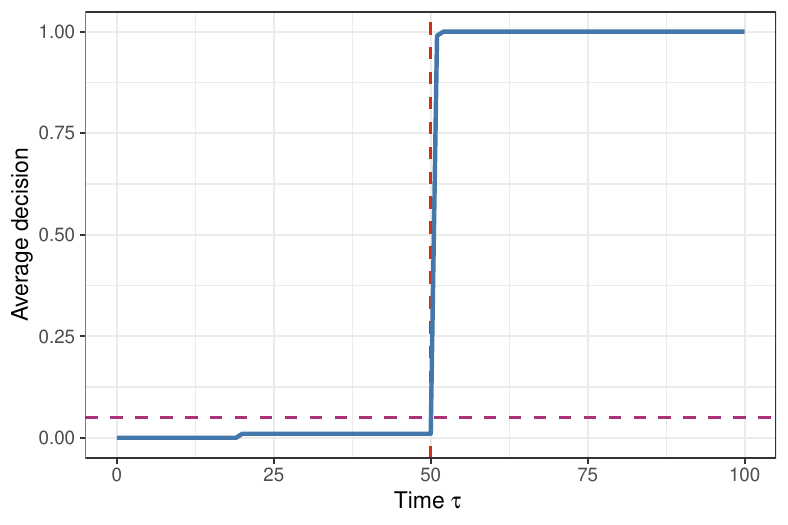}
\label{Laplace_change_Noise_Scale_Plot}
\end{subfigure}
\quad
\begin{subfigure}[c]{.49\linewidth}
\centering
\caption{\textbf{Laplace Mechanism - Change in Noise Type\,\,\bug}}
\includegraphics[width=0.95\linewidth,height=140pt]{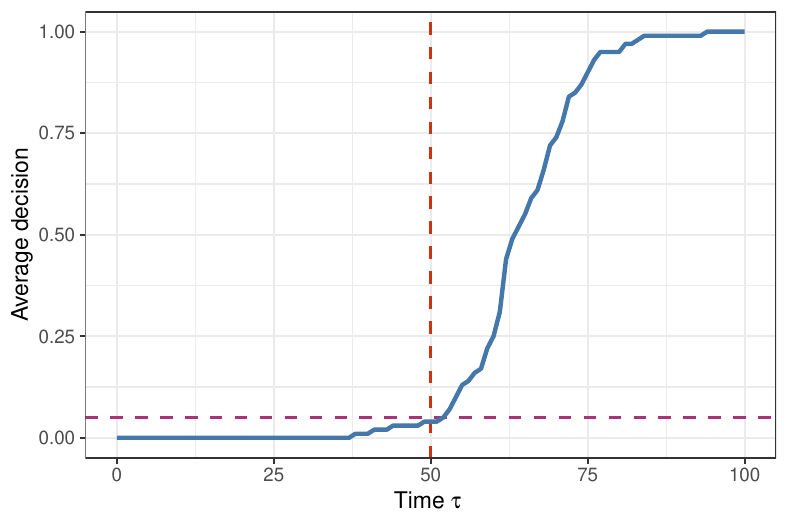}
\label{Laplace_change_Noise_Type_Plot}
\end{subfigure}

\begin{subfigure}[c]{.49\linewidth}
\centering
\caption{\textbf{Report Noisy Max - Change in Output Type}\,\,\bug}
\includegraphics[width=0.95\linewidth,height=140pt]{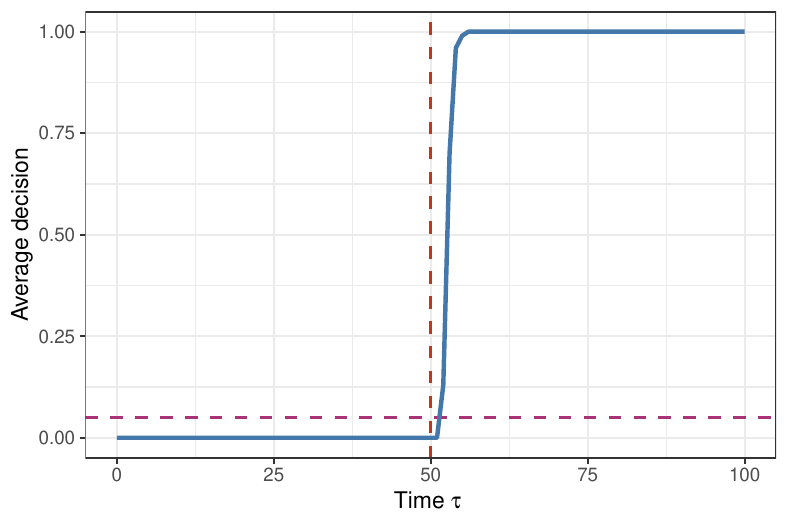}
\label{Report_Noisy_Max_change_Output_Type_Plot}
\end{subfigure}
\quad
\begin{subfigure}[c]{.49\linewidth}
\centering
\caption{\textbf{Change from SVT 2 to SVT 4\,\,\bug}}
\includegraphics[width=0.95\linewidth,height=140pt]{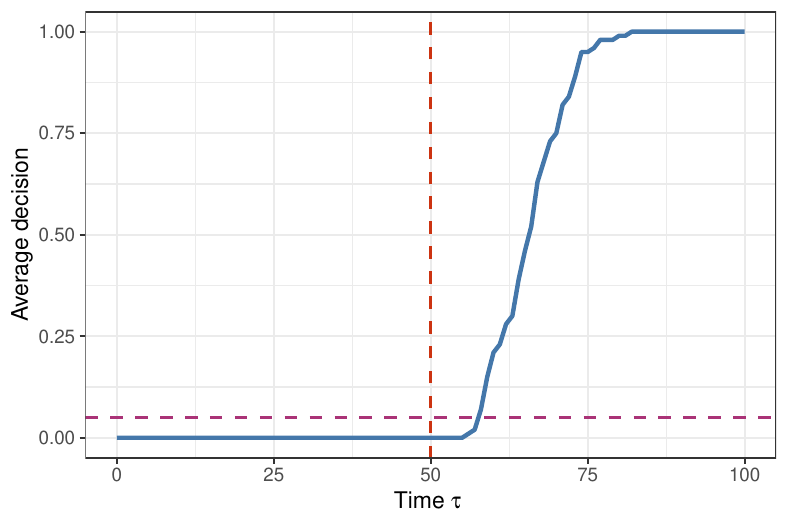}
\label{SVT_4_Plot}
\end{subfigure}

\begin{subfigure}[c]{.49\linewidth}
\centering
\caption{\textbf{Change from SVT 2 to SVT 5\,\,\bug}}
\includegraphics[width=0.95\linewidth,height=140pt]{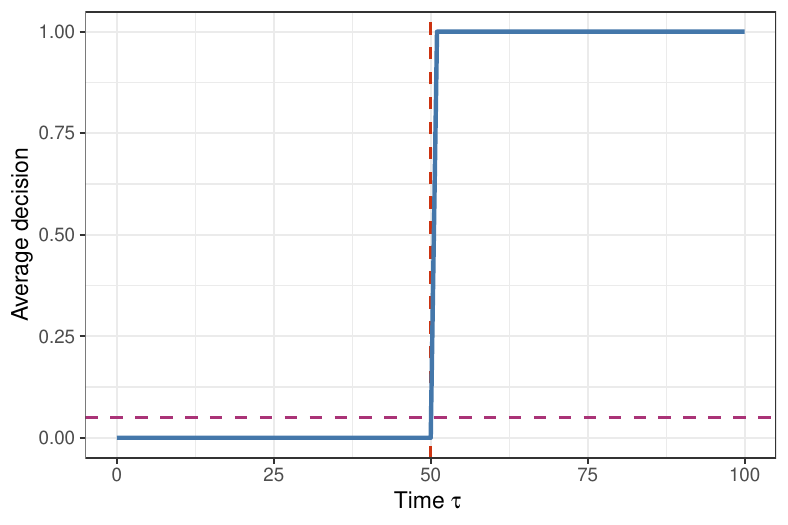}
\label{SVT_5_Plot}
\end{subfigure}
\quad
\begin{subfigure}[c]{.49\linewidth}
\centering
\caption{\textbf{Change from SVT 2 to SVT 6\,\,\bug}}
\includegraphics[width=0.95\linewidth,height=140pt]{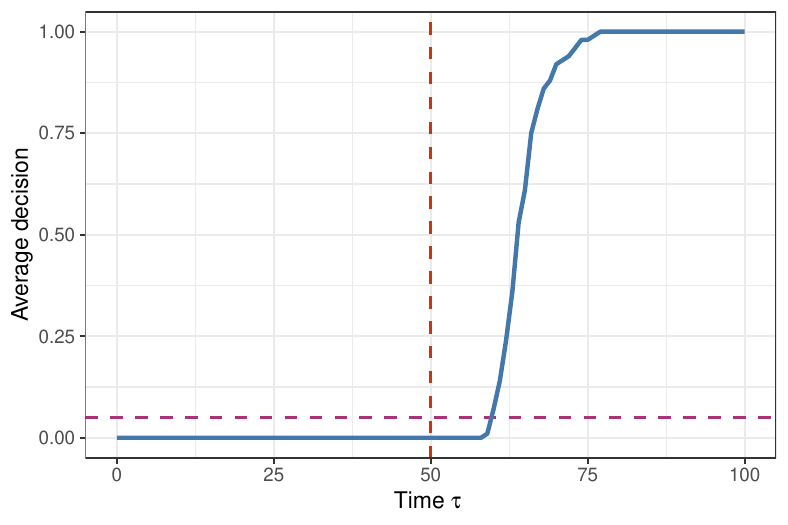}
\label{SVT_6_Plot}
\end{subfigure}

\begin{subfigure}[c]{.49\linewidth}
\centering
\caption{\textbf{Report Noisy Max - Change in Noise Type \,\,\nobug}}
\includegraphics[width=0.95\linewidth,height=140pt]{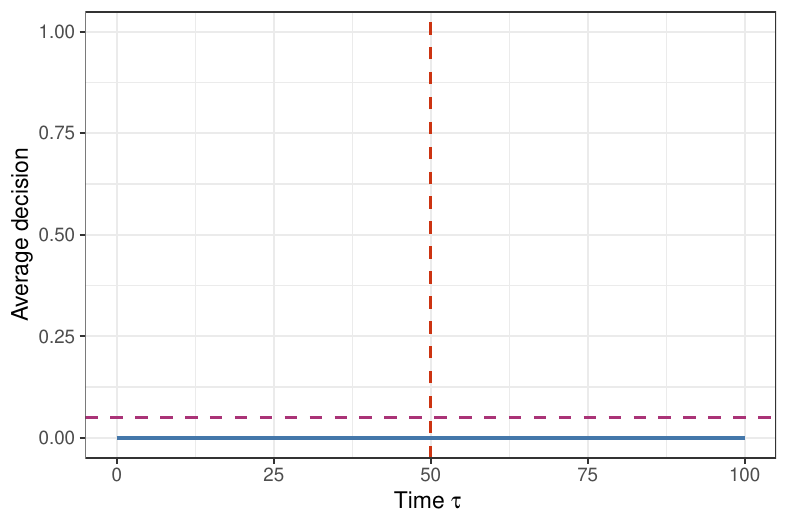}
\label{Report_Noisy_Max_change_Noise_Type_Plot}
\end{subfigure}
\quad
\begin{subfigure}[c]{.49\linewidth}
\centering
\caption{\textbf{Change from SVT 2 to SVT 1\,\,\nobug}}
\includegraphics[width=0.95\linewidth,height=140pt]{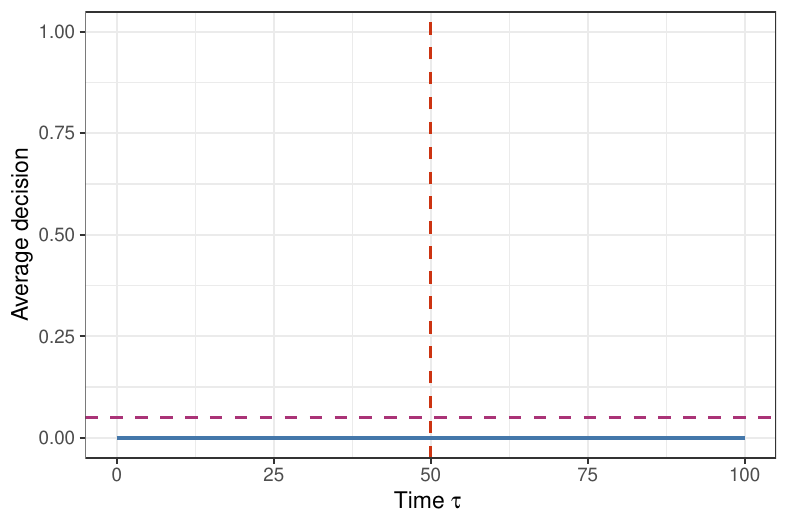}
\label{SVT_1_Plot}
\end{subfigure}
\caption{Average test decision of our detector. A modification happens at time $50$, where it either introduces a privacy violation
"\,\bug\," or not "\,\nobug\,".  \label{Figure_Algorithms}} 
\end{figure*}

\renewcommand{\arraystretch}{1.25}
\begin{table*}[t]
\begin{small}
\centering
\tabcolsep=0.12cm 
\begin{tabular}{@{}lllcccc@{}}
\toprule
\textbf{ } & \textbf{Algorithm} & \textbf{Modification} & \textbf{Database} $\boldsymbol{x}$ \textbf{or query $Q(x)$} & \textbf{Database} $\boldsymbol{x'}$ \textbf{or query $Q(x')$} & \textbf{Event} $\boldsymbol{\mathcal{E}}$  \\ \midrule
(a) & Laplace mechanism & Decrease noise scale & $(0,0,0,0,0,0,0,0,0,0)$ & $(1,0,0,0,0,0,0,0,0,0)$ & $(- \infty, 0]$ \\
(b)  & Laplace mechanism & Change Noise Type to Gaussian & $(0,0,0,0,0,0,0,0,0,0)$ & $(1,0,0,0,0,0,0,0,0,0)$ & $(- \infty, -1]$   \\
(c) & Report Noisy Max  & Report maximum noisy query & $(1,1,1,1,1)$ & $(2,2,2,2,2)$ & $(-\infty,2]$ \\
(d) & SVT 2  & Change to SVT 4 & $(0,0,0,0,0,1,1,1,1,1)$ & $(1,1,1,1,1,0,0,0,0,0)$ & \{(0,0,0,0,0,0,1)\}  \\
(e)  & SVT 2  & Change to SVT 5 & $(0,0,0,0,0,1,1,1,1,1)$ & $(1,1,1,1,1,0,0,0,0,0)$ & \{Q(x)\} \\
(f)  & SVT 2  & Change to SVT 6 & $(1,1,1,1,1,0,0,0,0,0)$ & $(0,0,0,0,0,1,1,1,1,1)$ & \{Q(x)\} \\
(g)  & Report Noisy Max  & Change Noise Type to Exponential & $(1,1,1,1,1)$ & $(2,2,2,2,2)$ & $\{3\}$  \\
(h)  & SVT 2  & Change to SVT 1 & $(0,0,0,0,0,1,1,1,1,1)$ & $(1,1,1,1,1,0,0,0,0,0)$ & \{(0,0,0,0,0,0,1)\}  \\ \bottomrule
\end{tabular}
\end{small}
\caption{\normalsize Summary of the algorithms investigated in our simulations.}
\label{tab:dp_scenarios}
\end{table*}

A flawed modification of SVT 2 encountered in the literature is SVT 4, where the calibration of noise in steps 2 and 4 of Algorithm \ref{Alg_SVT_2} is done without taking the bound $c$ into account, while the resampling of noise for the threshold in step 7 is omitted entirely. Other incorrect variants neglect to add noise to the queries $Q_i(x)$ (SVT 5) or fail to bound the number of noisy queries that lie above the noisy threshold (SVT 6). For our experiments, we undertake these adjustments to SVT 2 and test how fast our monitoring procedure detects these detrimental changes. We select $\varepsilon = 1$, $\Gamma = 1$ and $c =1$ in our experiments, and choose query patterns and events as in \cite{StatDP} for SVT 5 and SVT 6 (these can be found in Table \ref{tab:dp_scenarios}). A useful heuristic from \cite{StatDP} for choosing $\mathcal{E}$ is to choose the output we would obtain if we ran the algorithms without noise, which yields events of the form $\{ Q(x) \}$. For SVT 4, we choose the same query pattern as before, but use the approach described in \cite{Dette_2022} to identify an output value that is associated with high privacy leakage. Said output will then be a likely candidate for detecting a privacy violation. By doing so, we select the event $\mathcal{E} = \{(0,0,0,0,0,0,1)\}$. \\

\noindent \textbf{Benign Changes}
Not all modifications to an algorithm are harmful and we expect our monitoring procedure not to reject in these cases. For instance, applying Exponential noise $Exp(2 / \varepsilon)$ to our counting queries with sensitivity 1 should not affect the privacy of RNM \cite{StatDP}. Likewise, modifying SVT 2 in a manner that yields SVT 1 from \cite{Lyu_2017} should also have no negative impact on the privacy parameter $\varepsilon$. We apply our monitoring tool to these variants and choose the same parameters and query patterns as before. For $\mathcal{E}$, we again use the approach from \cite{Dette_2022} to find an output value that would indicate a privacy violation if it occurred. Our choice of databases and events can be found in more detail in Table \ref{tab:dp_scenarios}.

\subsection{Setup and Plots} \label{sec:4.2}
\noindent To illustrate our procedure, we consider various scenarios with both harmful and harmless modifications. Throughout these experiments, we have fixed the time horizon at $T=100$ and the sample size $n$ at each time point at $n=750$. Variations to the value of $n$ are explored below in Figure \ref{Figure_Different_n}.
In each scenario, a modification is introduced at time $\tau=50$.
We consider a total of $8$ scenarios, where a list  the deployed algorithms before and after the change is given in Table \ref{tab:dp_scenarios}. Notice that the letters (a)-(h) in that table correspond to the labels in Figure \ref{Figure_Algorithms}. 
Scenarios (a)-(f) correspond to harmful modifications (indicated by a \bug-symbol) and scenarios (g)-(h) to harmless modifications (indicated by a \nobug-symbol). 
In each scenario, we consider $100$ independent simulation runs to calculate $\{\widehat{D}(\tau): 1 \le \tau \le T\}$ 
and compare them to the threshold $q(\alpha)$ for $\alpha=0.05$ (details in Algorithm \ref{alg:dp_auditing}). If $\widehat{D}(\tau)>q(\alpha)$ we make the decision that "a privacy violation" has taken place and in Figure \ref{Figure_Algorithms}, we report at any time $\tau$ the proportion of runs where  a privacy violation has already been detected. The vertical line at $\tau=50$ in each plot marks the time at which the modification has actually taken place. The horizontal line at height $0.05$ in each plot indicates the user-defined FAR $\alpha$, which should not be exceeded in the absence of harmful changes. \\

\noindent \textbf{Results} Our experiments validate our procedure and illustrate key theoretical insights. First, we can see that in the absence of changes, the FAR is always upper bounded by $\alpha$. In Figure \ref{Figure_Algorithms} (a)-(f), the proportion of detected violations for $\tau<50$ is always lower than $0.05$ (the horizontal line). This means that before any harmful modification occurs, the FAR is $\le \alpha$. In scenarios (g)-(h) the change at time $50$ does not lead to a privacy violation and accordingly the proportion of detections stays below $\alpha$ over the entire time period.\\
Second, whenever the change has been harmful, that is, scenarios (a)-(f), it is detected eventually with high certainty. Indeed, in all of these scenarios the proportion of detected violations rises to $100\%$ in our experiments, showing that no violation is entirely overlooked. The speed of detection, indicated by the gradient of the curves, varies and is aligned with our intuition that stronger privacy violations can be detected faster. For instance, the adjustments undertaken to arrive at SVT 5 in scenario (e) are more severe than for other variants of SVT, where the changes made are more subtle. At the same time, we can observe that the privacy violation associated with SVT 5 is detected almost immediately, while smaller violations, as in scenario (d) and (f), take longer.

\begin{figure*}[t!]
\begin{subfigure}[c]{.49\linewidth}
\centering
\caption{\textbf{Panel for the Laplace mechanism} \,\,\,\bug}
\includegraphics[width=0.95\linewidth]{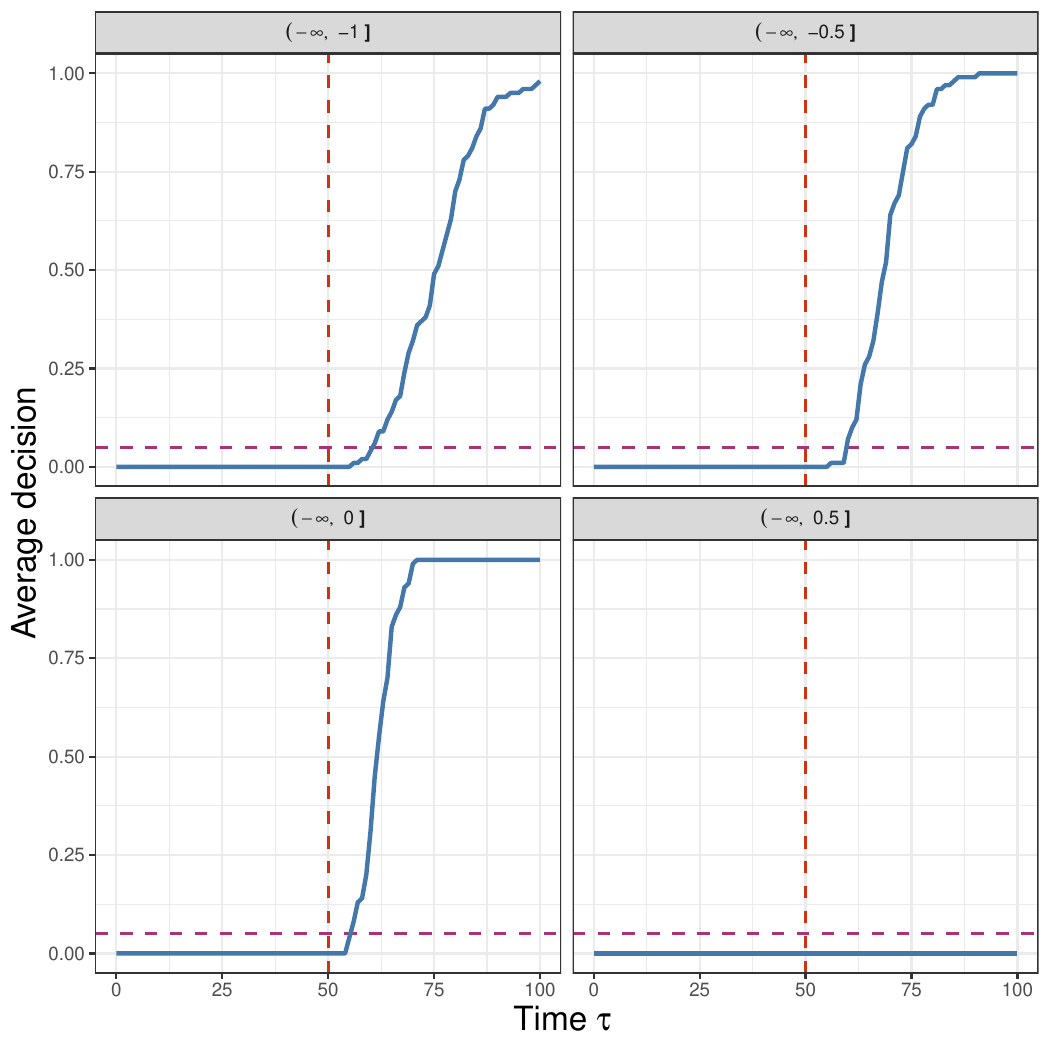}
\label{Panel_Plot}
\end{subfigure}
\quad
\begin{subfigure}[c]{.49\linewidth}
\centering
\caption{\textbf{Average Aggregate Decision}}
\includegraphics[width=0.95\linewidth]{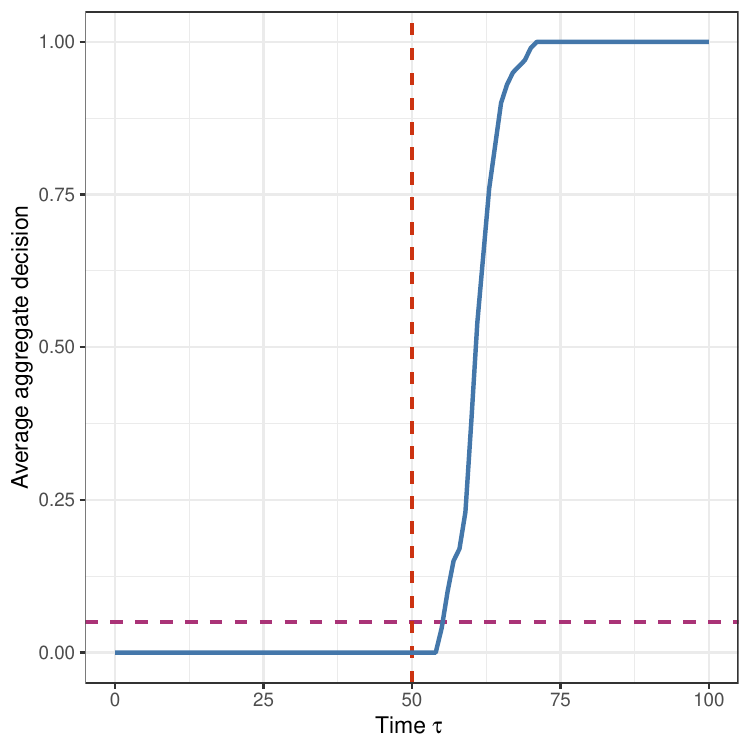}
\label{Aggregate_Decision_Plot}
\end{subfigure}

\caption{Panel and average aggregate decision   \label{Figure_Panel}} 
\end{figure*}

\subsection{Additional simulations}

\noindent \textbf{Multiple monitoring} In practice, it can be advisable to monitor an algorithm for several different configurations at the same time, since the choice of algorithm parameters, databases and events can influence how well privacy violations are detected. In that case, we need to aggregate the information from all monitoring runs into one decision for each time step $\tau$. We can do so by deciding that a privacy violations has occurred as soon as at least one of our monitoring runs detects a privacy violation. In order not to inflate the FAR, we apply a Bonferroni correction across all configurations.

For example, we can consider the Laplace mechanism from the previous subsection and modify its noise scale from $b = 1$ to $b = 0.9$ at time $\tau = 50$ (harmful modification). We choose the databases as before, but monitor the algorithm for four different, potentially vulnerable events. A Bonferroni correction implies that each monitoring scheme should then be run with level $\alpha = 0.0125$ to attain global FAR $0.05$. In Figure \ref{Figure_Panel}, we can see the effect of aggregating decisions across a panel of events. In the panel on the left-hand side, we see decisions based on a number of tail events $\mathcal{E}_a=(-\infty,a]$ with $a\in \{-1,-0.5,0,0.5\}$. Whenever any one of the individual detectors decides that a privacy violation has taken place, the aggregate detector (displayed on the right) also decides that a violation has taken place. Consequently, the curve of the aggregate detector rises faster than any of the individual curves. We make two observations: First, even if one of the individual detectors is weak, the aggregate detector can still be strong. For example, the detector based on the event $\mathcal{E}_{0.5}$ barely detects any violations at all. However, the aggregate detector, drawing on information from the other events, still manages to rapidly identify the harmful change with small delay. Second, and more subtle, the events $\mathcal{E}_{a}$ for $a=-1, -0.5, 0$ all lead to the same theoretical $\varepsilon$. Nevertheless, we see that $\mathcal{E}_{0}$ produces the fastest increasing curve and dominates the aggregate detector.\\

\begin{figure*}[t!]
\begin{subfigure}[c]{.49\linewidth}
\centering
\caption{\textbf{Laplace Mechanism - Change in Noise Type\,\,\bug}}
\includegraphics[width=0.95\linewidth]{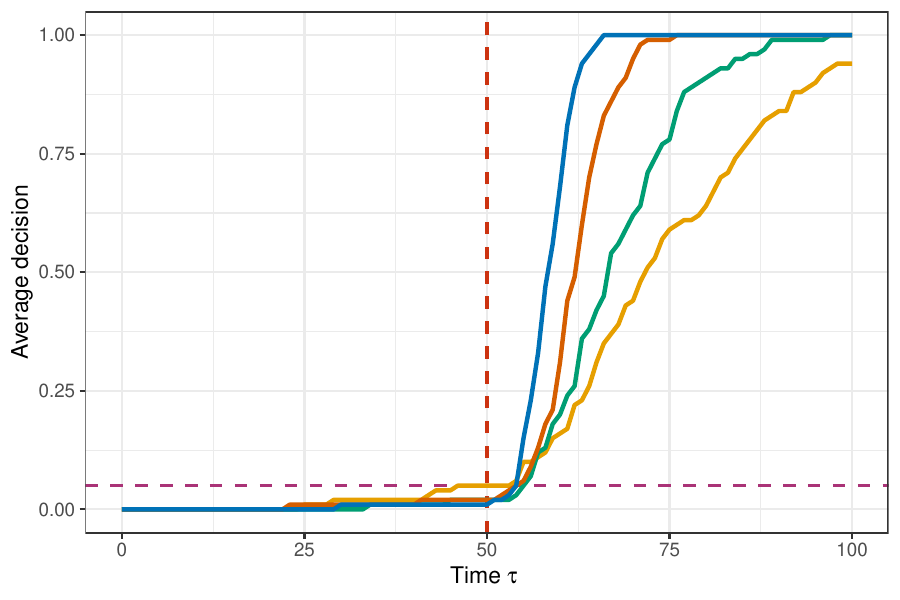}
\label{Laplace_Different_n}
\end{subfigure}
\quad
\begin{subfigure}[c]{.49\linewidth}
\centering
\caption{\textbf{Change from SVT 2 to SVT 4\,\,\bug}}
\includegraphics[width=0.95\linewidth]{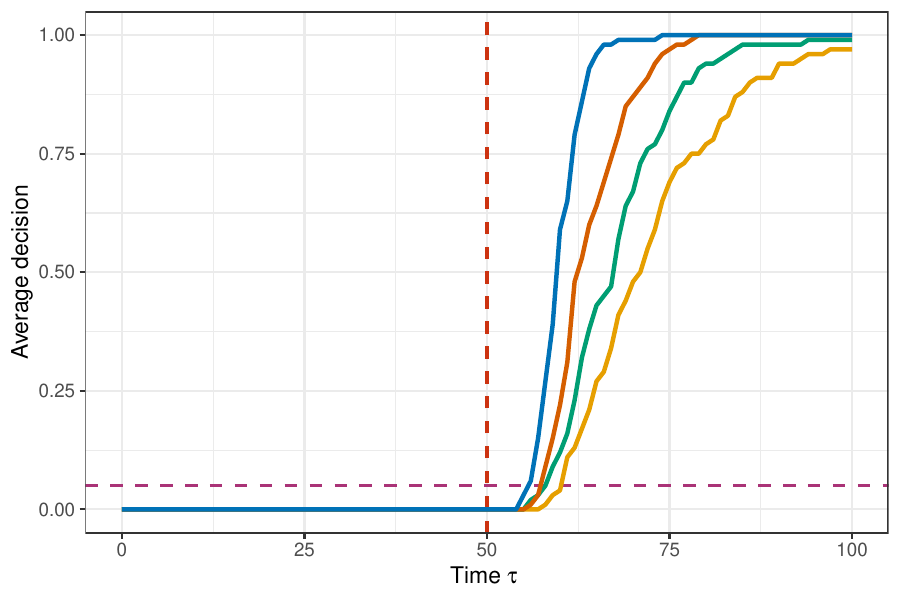}
\label{SVT_4_Different_n}
\end{subfigure}

\caption{Scenario (b) and (d) from Figure \ref{Figure_Algorithms} for $n = 200$ (yellow), $500$ (green), $1000$ (red), $2000$ (blue)  \label{Figure_Different_n}} 
\end{figure*}

\noindent \textbf{The impact of $n$} Finally, we want to study the effect of different values of the sample size $n$, available at each time point. For this purpose, we consider scenarios (b) and (d) from Section \ref{sec:4.2}. All experiment specifications are kept the same, except, that we repeat the experiments with different sizes of $n$, namely $n=200, 500, 1000,2000$. Results are displayed in Figure \ref{Figure_Different_n}. First, we notice that in all cases before the change, the FAR is below $0.05$ (the vertical line) for all choices of $n$ and for both algorithms, which is in line with the guarantees of Theorem \ref{theo:main}. Theorem \ref{e:theo:delay} in turn states that $n$ should have an influence on the detection delay, where larger $n$ should lead to a faster detection of the change. We see this in effect for both algorithms in Figure \ref{Figure_Different_n}, where consistently larger $n$ are associated with a faster increase of alarms after the change has occurred (i.e. faster detection).

\section{Related work} \label{Section_Related_Work}

\noindent Our work is conceptually closest to previous approaches that audit algorithms with the help of tuples $(x,x',\mathcal{E})$ \cite{StatDP, DP-Finder, DP-Sniper}. A comparison to these works also underlines the benefits of developing a method that is specifically designed for the monitoring task over extended periods of time. For instance, the auditor in \cite{StatDP} and \cite{DP-Sniper} use up to $n = 5\times 10^5$ and $n = 10.7 \times 10^6$ samples respectively to audit an algorithm once, and repeating this procedure over time would lead to high sampling costs. Our method works with $n = 750$ samples at each time $t$, a fraction of the samples used in these previous works.

Many related works, however, rely on other elements than the tuples $(x,x', \mathcal{E})$ to audit DP. This is especially true for the various variants of DP, where privacy loss can oftentimes be expressed with the help of divergences. Accordingly, a number of approaches are built on these and other statistical distances rather than the tuples used in our approach \cite{Liu_2019, Kutta_2024, Kong_2024, Lu_2024}. And many auditing approaches that are tailored to DP-SGD, an important DP mechanism for privacy-preserving machine learning, make use of $f$-differential privacy \cite{Nasr_2023, Annamalai_2024,Annamalai_2024_b}, another variant of DP that works without the aforementioned tuples. We believe that an interesting direction for future work would be to design a monitoring procedure that works for these and other approaches.

\noindent \textbf{Sequential monitoring}  
In this paper, we consider sequential monitoring for changes in differential privacy (DP) while controlling the false alarm rate (FAR), as detailed in Section \ref{Section_Monitoring}. The paradigm of monitoring with FAR control was introduced in the seminal work of \cite{chu:stinchcombe:white:1996} and has since become a cornerstone of statistical real-time data analysis. For a recent overview, see \cite{aue:kirch:2024}.  
Alternative approaches to real-time analysis include methods developed in the field of "Stochastic Process Control." For an overview, we refer to the monograph of \cite{qiu2014}. Statistical monitoring has been developed to various data types, including real-valued data \cite{horvath:kokoszka:huskova:steinebach:2003}, panel data \cite{horvath:kokoszka:wang:2021}, and more general data objects \cite{kutta:jach:kokoszka:2025}. The challenge of achieving short detection delays, which is also a focus of this paper, has been addressed by \cite{aue:horvath:2004}, \cite{yu:padilla:wang:rinaldo:2023}, and \cite{kutta:doernemann:2025}.  \\
Like the works cited above, this paper employs fundamental results from stochastic process theory to control the FAR. However, our analysis differs in its objective. Classical statistical monitoring typically aims to detect distributional changes in a data stream compared to a baseline period. For such methods, any distributional change is relevant. In contrast, our approach is tailored to applications in DP, where algorithmic modifications are of interest only if they are harmful (i.e., introduce privacy breaches). Furthermore, our framework does not rely on a "baseline period". Rather, we  monitor adherence to a fixed $\varepsilon$-DP threshold, ensuring that it is not violated.

\section{Conclusion} \label{Section_Conclusion}

\noindent We introduced a new problem: the continuous monitoring of a randomized algorithm for DP breaches. This problem extends the scope of the DP auditing literature, which has traditionally treated auditing as a one-time task. For algorithms undergoing changes during development or deployment, consistent monitoring might be warranted to ensure sustained privacy.
To address this, we proposed a novel method for aggregating a sequence of weak auditors into a reliable detector that provably controls false alarms. The aggregation mechanism employs a weighted majority vote over the weak auditors, assigning the greatest influence to the most recent observations.
Our experiments demonstrate that this approach reliably and efficiently detects harmful changes across a variety of standard DP algorithms, while maintaining moderate sampling requirements for each screening.

\appendix
\section*{Acknowledgements}
This work was funded by the Deutsche Forschungsgemeinschaft (DFG, German Research Foundation) under Germany's Excellence Strategy - EXC 2092 CASA - 390781972. Tim Kutta's work has been partially funded by AUFF grants 47331 and 47222.

\bibliographystyle{IEEEtran}
\bibliography{references}

\section*{Appendix}

\appendix

The appendix is dedicated to the proofs of our theoretical results and to giving additional technical details. Throughout the appendix we will use the notation $\lesssim $ when comparing two sequences of positive numbers $(a_n)_n $ and $(b_n)_n$. More precisely, we write $a_n \lesssim b_n$ if there exists a constant $C>0$ (independent of $n$) such that $a_n \le C b_n$ for all $n \ge 1$. For a sequence of random variables $(Z_n)_n$ we denote weak convergence to a limit $Z$ by $Z_n \overset{d}{\to}Z$.

\section{Proofs for the naive method} \label{sec:naive}

We prove the normal approximation in \eqref{e:pt/s} for fixed output distribution at $t$ and for $n \to \infty$. For this convergence to hold it is necessary that $\hat p_t$ does not have a degenerate distribution which means that its variance is $>0$ (i.e. $\sigma_t^2>0$ defined in \eqref{e:var:pt} below).
If \eqref{e:pt/s} holds, or more mathematically, if the weak convergence
\begin{align} \label{e:pt/s2}
    \frac{\hat p_t-p_t}{\hat \sigma_t} \overset{d}{\to}\mathcal{N}(0,1)
\end{align}
holds it follows directly that
\[
\lim_{n \to \infty}\mathbb{P}\big( p_t \in \mathcal{C}_t(\alpha)\big)=\alpha.
\]
To show \eqref{e:pt/s2}, we first compute the variance of $\hat p_t$:
\begin{align} \label{e:var:pt}
    \sigma_t^2:=&n\mathbb{V}ar(\hat p_t) =  \frac{\mathbb{V}ar(n_t^X) + e^{2\epsilon}\, \mathbb{V}ar(n_t^Y) }{n} \\
    =& q_t^X(1-q_t^X) +e^{2\epsilon}q_t^Y(1-q_t^Y)  . \nonumber
\end{align}
Here, we have defined the success probabilities of the binomials involved as
\begin{align*}
    q_t^X := \mathbb{P}(A(x) \in \mathcal{E}), \quad  q_t^Y := \mathbb{P}(A(x') \in \mathcal{E}).
\end{align*}
For the statement in \eqref{e:pt/s2}, it suffices to show that $n\hat \sigma_t^2$ converges to $\sigma_t^2$ in probability. We demonstrate a stronger statement that has additional use when studying the aggregated auditor below. Namely we show
\begin{align} \label{e:var:con}
    \max_t |\sigma_t^2-n\hat \sigma_t^2|=o_P(1). 
\end{align}
Here "$o_P(1)$" means convergence to $0$ stochastically. We know that there exists an even natural number $M$ such that $T \lesssim n^{M/4}$ (Assumption \ref{ass:1}). Then we have, using Lemma 2.2.2. in \cite{vaart:wellner:1996}, that
\begin{align*}
    &\mathbb{E}\max_t |\sigma_t^2-n\hat \sigma_t^2| \lesssim T^{1/M} \{\mathbb{E}\max_t |\sigma_t^2-n\hat \sigma_t^2|^M\}^{1/M} \\
    \lesssim & n^{1/4} \{\mathbb{E}\max_t |\sigma_t^2-n\hat \sigma_t^2|^M\}^{1/M}. 
\end{align*}
The final step of proving 
\begin{align} \label{e:var}
\{\mathbb{E}\max_t |\sigma_t^2-n\hat \sigma_t^2|^M\}^{1/M} = o(n^{-1/4} )  
\end{align}
follows by standard techniques (using uniform boundedness of moments for standardized binomial random variables) and is hence omitted. \\
Next, it is clear (by linearity of the expectation) that $\mathbb{E}\hat p_t =p_t$ and finally, using the classical central limit theorem  for binomial random variables that
\[
n^{1/2}(\hat p_t-\mathbb{E}\hat p_t) =n^{1/2}(\hat p_t- p_t) \to \mathcal{N}(0,\sigma_t^2).
\]
Now, combining this with the fact that $n\hat \sigma_t^2-\sigma_t^2=o_P(1)$ and using Slutsky's theorem yields \eqref{e:pt/s2}.

\section{Stochastic processes on Hölder spaces} \label{sec:stoch}

The mathematics of this paper are rooted in "stochastic process theory" a branch of probability theory. More precisely, we study stochastic processes that are Hölder continuous. We now give a rough outline of the involved objects and refer for a more detailed introduction to the work of \cite{ravckauskas:suquet:2009}. \\
We consider a continuous function $f: [0,1] \to \mathbb{R}$. For a parameter $\beta \in [0,1/2)$ we define its Hölder norm
\[
\|f\|_\beta := \sup_{t \in [0,1]} |f(t)| + \sup_{0 \le s < t \le 1} \frac{|f(t)-f(s)|}{|t-s|^\beta}.
\]
Moreover, we say that $f$ has a vanishing Hölder modulus of continuity if
\[
\lim_{\delta \downarrow 0} 
\sup_{\substack{0 \leq s < t \leq 1 \\ |s-t| \leq \delta}} 
\frac{|f(t) - f(s)|}{|t-s|^\beta} = 0.
\]
The collection of all continuous functions $f$ that have finite $\beta$-Hölder norm $\|f\|_\beta<\infty$ and have vanishing Hölder modulus of continuity make up the Hölder space $\mathcal{H}_\beta[0,1]$. Sometimes we will define the subspace $\mathcal{H}_\beta[0,\theta]$ which consists of the restrictions on that interval (the Hölder norm is defined in analogy). The Hölder space is for any $\beta \in [0,1/2)$ a separable Banach space and for $\beta=0$ it is identical to the continuous functions.\\
A stochastic process on $\mathcal{H}_\beta[0,1]$  (also: with sample paths in $\mathcal{H}_\beta[0,1]$) is a measurable random variable that takes values in $\mathcal{H}_\beta[0,1]$ . One example is the sBm. For a sequence of stochastic processes $(P_n)_n$ in $\mathcal{H}_\beta[0,1]$ to converge to a limit $P$ in $\mathcal{H}_\beta[0,1]$, this means that $P_n$ converges weakly w.r.t. to the topology on $\mathcal{H}_\beta[0,1]$ created by the Hölder norm. To understand in detail what this means we refer to Section 2 in  \cite{ravckauskas:suquet:2009}, where also additional references are given.

\section{Preliminary results on stochastic processes} \label{sec:glue}

A key technique for our proofs is reducing the convergence of a (more complicated) stochastic process $P_n$, to convergence of its simpler segments. For this purpose we introduce the following "gluing operation", where a stochastic process is cut into separate segments that converge and finally put together again to show convergence of the entire process.

\textbf{Gluing operations for stochastic processes} Suppose we have a stochastic process (random function) $P_n(x), 0 \le x \le 1$ indexed in $n$. Moreover, suppose that there exists a fixed number of points $\theta_0:=0<\theta_1<\cdots<\theta_K<1$ such that for each $k=0,...,K-1$ we define
\[
G_n^{(k)}(x):=P_n(x+\theta_k)-P_n(\theta_k), \,\, 0 \le x \le \theta_{k+1}-\theta_k.
\]
Further suppose that each $G_n^{(k)}$ converges weakly to an sBm $B^{(k)}$ in the Hölder space $\mathcal{H}^\beta[0,\theta_{k+1}-\theta_k]$ and the sBms are independent across $k$. Then, $P_n(x), 0 \le x \le 1$ converges weakly to $B(x), 0 \le x \le 1$ in the Hölder space $\mathcal{H}^\beta[0,1]$, where $B$ is an sBm. 

\textbf{Proof of gluing statement} The Hölder space $\mathcal{H}^\beta[0,\theta_{k+1}-\theta_k]$ is complete and separable, and thus the same is true of the product space
\[
\prod_{k=0}^{K-1}\mathcal{H}^\beta[0,\theta_{k+1}-\theta_k]
\]
equipped with the maximum metric. 
Using Skorohod's theorem (see \cite{billingsley:1999}, Theorem 6.7), we can switch to a probability space where $G_n^{(k)}$ converges almost surely to $B^{(k)}$ in the Hölder metric. Recall that $P_n$ (which is the glued-together version of  $G_n^{(k)}$) is defined for $\theta_k \le x \le \theta_{k+1}$ as
\[
P_n(x) = G_n^{(k)}(x-\theta_k) +  \sum_{\ell=0}^{k-1} G_n^{(\ell)}(\theta_{\ell+1}-\theta_\ell).
\]
This object now converges 
converges almost surely to $B$, which is the glued together sBm, defined in $\theta_k \le x \le \theta_{k+1}$ as
\[
B(x) = B^{(k)}(x-\theta_k) +  \sum_{\ell=0}^{k-1} B^{(\ell)}(\theta_{\ell+1}-\theta_\ell).
\]
It is trivial to check that $B$ is an sBm and we omit the proof. To see the almost sure convergence of $P_n$, fix the outcome on the probability space such that all $G_n^{(k)}$ converge to $B^{(k)}$. Then, we have 
\[
\|B-P_n\|_\beta \le \sum_k \|B^{(k)}-G_n^{(k)}\|_\beta \to 0.
\]

\section{Proof of Theorem \ref{theo:main}}

The proof of this Theorem involves a number of stochastic processes and sums that are (minor) modifications of each other. Since it is easy to lose track of what a specific modification means and why it is made, we have added an overview in the below  \hyperref[box:minimalist]{Infobox} near the end of this section.

\textbf{Preliminary considerations} In this proof we denote $A^{(1)}:= A_1$, $A^{(2)} := A_{t_1},\cdots,A^{(K+1)}:=A_{t_K}$ (to refer to the algorithm in any of the periods were it does not undergo a change. We also define the objects
\[
q^{(k),X} := \mathbb{P}(A^{(k)}(x) \in \mathcal{E}), \qquad q^{(k),Y} :=\mathbb{P}(A^{(k)}(x') \in \mathcal{E}).
\]
Now, we will assume in this proof that for $k=1,...,K+1$ it holds that
\[
q^{(k),X} +q^{(k),Y}  >0.
\]
The (easy) modification where this value is $=0$ for some $k$ is ignored for the sake of compactness in this and the following proofs. We will furthermore only consider the case in this proof where $T \to \infty$ as $n \to \infty$. The case where $T$ is fixed and $n \to \infty$ is again much easier.

\textbf{Gluing and weak convergence} For the proof of the theorem, we consider the stochastic process 
\[
P_n(x) := \sum_{i=1}^{\lfloor x T \rfloor} \frac{\hat p_t-p_t}{\hat \sigma_t\sqrt{T}} +\big[x-(\lfloor T x \rfloor/T )\big] \frac{\hat p_{\lceil T x \rceil}-p_{\lceil T x \rceil}}{\hat \sigma_{\lceil T x \rceil}\sqrt{T}}.
\]
This process depends on $n$ by the definition of $\hat p_t, \hat \sigma_t$ and $T$. The main step of this proof is now showing weak convergence of $P_n$ to an sBm on the Hölder space $\mathcal{H}_\beta[0,1]$, i.e. 
\[
\{P_n(x): 0 \le x \le 1\} \overset{d}{\to} \{B(x): 0 \le x \le 1\}.
\]
According to our considerations in Section \ref{sec:glue}, it suffices to show separately that the processes $G_n^{(k)}$ (defined there) converge individually to an sBm on the subintervals $[0,   \theta_{k+1}-\theta_k]$. Each subinterval can be treated in the same way and hence, we focus on the one with $k=0$, on which $G_n^{(0)}=P_n$.
Thus, we now need to show 
\[
\{P_n(x): 0 \le x \le \theta_1\} \overset{d}{\to} \{B(x): 0 \le x \le \theta_1\},
\]
to obtain convergence.\footnote{A technicality that we do not discuss in detail is the independence of the limits $B^{(k)}$ of $G_n^{(k)}$ for $k=0,...,K-1$, required for the gluing result in Section \ref{sec:glue}. We just summarize the main idea here: The processes $G_n^{(k)}$ are already independent when $(T\theta_k)_k$ are all natural numbers. In general 
the processes $G_n^{(k)}$ are "almost independent", in the sense that they are independent except for a negligible error of size $\mathcal{O}_P(T^{-1/2})$ that asymptotically vanishes. Thus their limits are completely independent.} From now on, we will consider $P_n$ and all other following processes restricted to the Hölder space $\mathcal{H}^\beta[0,\theta_1]$.
We define
\begin{align*}
\widetilde{P}_n(x) = &P_n(x) \mathbb{I}\{0 \le x \le \lfloor T \theta_1 \rfloor/T\} \\
+ & P_n(\lfloor T \theta_1\rfloor /T) \mathbb{I}\{x>\lfloor T \theta_1 \rfloor/T\} 
\end{align*}
and elementary calculations show that
\[
\|\widetilde{P}_n-P_n\|_\beta=o_P(1).
\]
What $\widetilde{P}_n$ does is that it only uses data from before the modification time $\theta_1$ (the very short period $\lfloor T \theta_1 \rfloor/T<x \le T \theta_1$ it then remains constant).
Given the asymptotic identity of $P_n, \widetilde{P}_n$,  it suffices to show the weak convergence for $\widetilde{P}_n$ to an sBm $B$.

\textbf{Weak convergence of $\widetilde{P}_n$} 
 At the beginning of this proof, we said that $\hat p_t$ has a positive variance for any $t$ (see eq. \eqref{e:var:pt}). Recall that the true variance of $\hat p_t$ is $\sigma_t^2/n$. Since $\widetilde{P}_n$only involves information of time points $1 \le t \le t_1 = \lfloor T\theta_1 \rfloor$, all of the $\hat p_t$ are i.i.d. and have the same positive variance, which is $\sigma_1^2/n$.
Using the finding in \eqref{e:var:con}, it follows that with probability going to $1$ it holds that (for all $t$ simultaneously)
\[
\frac{\hat p_t-p_t}{\hat \sigma_t} = \frac{\hat p_t-p_t}{\hat \sigma_t} \mathbb{I}\{n^{1/2}\hat \sigma_t \ge \sigma_1/2\}.
\]
Defining for $t=1,...,\lfloor T\theta_1\rfloor$
\[
Z_t:= \frac{\hat p_t-p_t}{\hat \sigma_t} \mathbb{I}\{n^{1/2}\hat \sigma_t \ge \sigma_1/2\}
\]
and $Z_{\lfloor T\theta_1\rfloor+1} $ as in independent copy of these (i.i.d.) variables, we obtains that 
\begin{align*}
    \widetilde{P}_n(x) = P_n'(x)+o_P(1),
\end{align*}
where the remainder is  vanishing in the Hölder norm. Here 
\begin{align*}
   P_n'(x) :=&\sum_{i=1}^{\lfloor x T \rfloor} \frac{Z_i}{\sqrt{T}} +\big[x-(\lfloor T x \rfloor/T )\big] \frac{Z_{\lceil T x \rceil}}{\sqrt{T}}.
\end{align*}
Now, we may show weak convergence of $ P_n'$ on the Hölder space to an sBm.
According to Section 2.1 in \cite{ravckauskas:suquet:2009} (observations a) and b) on p. 265) we need to show weak convergence of the marginal distributions of $  P_n'$ and tightness in the Hölder space. If we can show marginal convergence, tightness already follows from criterion ii) on p. 266 in  \cite{ravckauskas:suquet:2009}. This criterion can be validated following exactly the steps in the proof of Lemma 3.3 in \cite{kutta:doernemann:2025}. This leaves us with proving marginal convergence. In principle the proof needs to be conducted using multivariate marginals, but for illustration we will confine ourselves to showing that $ P_n'(\theta_1)$ converges to a standard normal.\\
We have that 
\[
P_n'(\theta_1)= \sum_{i=1}^{\lfloor  T \theta_1\rfloor} \frac{Z_i}{\sqrt{T}} +o_P(1)=:C_1 +o_P(1). 
\]
Here $C_1$ is defined in the obvious way.
Next, we introduce the variables 
\[
\check{Z}_i:=\frac{\hat p_t-p_t}{ \sigma_t/n^{1/2}} \mathbb{I}\{n^{1/2}\hat \sigma_t \ge \sigma_1/2\}
\]
and therewith 
\[
C_2 := \sum_{i=1}^{\lfloor  T \theta_1\rfloor} \frac{\check{Z_i}}{\sqrt{T}}.
\]
We now demonstrate that
$C_1-C_2=o_P(1)$. Notice therefore that
\begin{align*}
   & \mathbb{E}[C_1-C_2]^2 \\
   =&  \frac{1}{T}\sum_t \mathbb{E}\bigg[ \Big(\frac{(\hat p_t-p_t)(\hat \sigma_t-\sigma_t/n^{1/2})}{ (\sigma_t/n^{1/2}) \hat \sigma_t} \Big)\mathbb{I}\{n^{1/2}\hat \sigma_t  \ge \sigma_1/2\}\bigg]^2\\
    =&  \frac{1}{T}\sum_t \mathbb{E}\bigg[ \Big(\frac{n^{1/2}(\hat p_t-p_t)(n^{1/2}\hat \sigma_t-\sigma_t)}{ \sigma_t (n^{1/2}\hat \sigma_t)} \Big)\\
    & \cdot \mathbb{I}\{n^{1/2}\hat \sigma_t  \ge \sigma_1/2\}\bigg]^2\\
    \lesssim &  \mathbb{E}[n^{1/2}(\hat p_t-p_t)(n^{1/2}\hat \sigma_t-\sigma_t)]^2 \\
    \le &  \{\mathbb{E}(n^{1/2}[\hat p_t-p_t])^4\}^{1/2}\{\mathbb{E}(n^{1/2}\hat \sigma_t- \sigma_t)^4\}^{1/2}.
\end{align*}
It is elementary to show that
\[
\{\mathbb{E}(n^{1/2}[\hat p_t-p_t])^4\}^{1/2} =\mathcal{O}(1).
\]
Using \eqref{e:var} we can conclude that 
\[
\{\mathbb{E}(n^{1/2}\hat \sigma_t-\sigma_t)^4\}^{1/2}=o(1)
\]
and thus the claim that $C_1-C_2=o_P(1)$ follows directly. Next, we define 
\[
C_3 := \sum_{i=1}^{\lfloor  T \theta_1\rfloor} \frac{\bar{Z_i}}{\sqrt{T}},
\]
where
\[
\bar{Z}_i:=\frac{\hat p_t-p_t}{ \sigma_t/n^{1/2}}.
\]
Using again the fact that $n^{1/2}\hat \sigma_t  \ge \sigma_1/2$ holds asymptotically with probability going to $1$ and simultaneously for all $t$, it follows that $C_2=C_3$ with probability going to $1$. Now, $C_3$ is easy to investigate. 
Asymptotic normality for $C_3$ can now be established using the Lindeberg central limit theorem. Notice that the variance of $\bar Z_t$ is by construction $1/T$. It follows that
\[
C_3 \overset{d}{\to} \mathcal{N}(0,\theta_1).
\]
This finishes the proof of the fact that $P_n$ converges to an sBm on the Hölder space. The next Lemma shows weak convergence of a version of $\widehat{D}$.

\begin{tcolorbox}[
    colback=white,      % White background
    colframe=gray!50,   % Border color
    sharp corners,      % Remove rounded corners
    boxrule=0.5mm,      % Border thickness
    width=0.5\textwidth, % Box width
    left=2mm,           % Left padding
    right=2mm,          % Right padding
    top=2mm,            % Top padding
    bottom=2mm          % Bottom padding
]
\textbf{Summary of stochastic processes and their use in the proofs\\}
\begin{itemize}
    \item[1)] ($B^{(k)}$), $B$: (segment of) sBm; it occurs as limiting distribution on the Hölder space $\mathcal{H}_\beta$ of $P_n$.
    \item[2)] $P_n$: partial sum process of the statistics $(\hat p_t-p_t)/\hat \sigma_t$; it is possible to write the detector $\widehat{D}(\tau)$ using $P_n$ and thus studying the convergence of $P_n$ is integral to our proof.
    \item[3)] $\widetilde{P}_n$: We study the convergence of $P_n$ in the proof of Theorem \ref{theo:main} restricted to the interval $[0,\theta_1]$; but $\lfloor T \theta_1\rfloor/T$ may be smaller than $\theta_1$ and thus $P_n$ uses information from after the change point $\theta_1$; to avoid this complication we create  $\widetilde{P}_n$ which only uses data from times $1 \le t\le \lfloor T \theta_1\rfloor$.
    \item[4)] $P_n'$: the individual terms in $\widetilde{P}_n$ are divided by $\hat \sigma_t$; but $\hat \sigma_t$ can be close to $0$ (with small probability); these cases make it hard to study weak convergence; so we move to a process where  $\hat \sigma_t$ (times $n^{1/2}$) is bounded away from $0$, namely $P_n'$.
  \item[5)] $C_1, C_2,C_3$: $C_1=P_n'(\theta_1)$, i.e. a sum of i.i.d. random variables; to more conveniently apply a CLT we replace $C_1$ by $C_3$ which consists of i.i.d. variables that all have variance $1/T$; $C_2$ is used for an intermediary replacement step.
\end{itemize}
\label{box:minimalist} 
\end{tcolorbox}

\begin{lemma} \label{lem:_help}
    Suppose that on the Hölder space $\mathcal{H}^\beta[0,1]$ the weak convergence 
    \[
\{P_n(x): 0 \le x \le 1\} \overset{d}{\to} \{B(x): 0 \le x \le 1\}.
\]
holds, where $B$ is an sBm.
Define 
\[
 \widetilde{D}(\tau):=  \max_{0 \le \ell < \tau} \bigg[\frac{1}{(\ell+1)^\beta \,\cdot\,T^{1/2-\beta}}\sum_{t=\tau-\ell}^\tau \frac{\hat p_t-p_t}{\hat \sigma_t}\bigg].
 \]
Then, it also follows that
\[
\max_{1 \le \tau \le T}  \widetilde{D}(\tau) \overset{d}{\to}\sup_{0\le x<y \le 1} \frac{B(y)-B(x)}{(y-x)^{\beta}}=D.
\]
\end{lemma}
For the proof, the reader first needs to observe that by some additional arguments
\[
\max_{1 \le \tau \le T}  \widetilde{D}(\tau)  = \sup_{0\le x<y \le 1} \frac{P_n(y)-P_n(x)}{(y-x)^{\beta}}+o_P(1).
\]
Now, the lemma is a direct consequence of the continuous mapping theorem, combined with the fact that the mapping 
\[
\mathcal{H}^\beta[0,1] \ni f \mapsto \sup_{0\le x<y \le 1} \frac{f(y)-f(x)}{(y-x)^{\beta}} \in \mathbb{R}
\]
is continuous. Because then we have 
\begin{align*}
    \sup_{0\le x<y \le 1} \frac{P_n(y)-P_n(x)}{(y-x)^{\beta}}\overset{d}{\to}\sup_{0\le x<y \le 1} \frac{B(y)-B(x)}{(y-x)^{\beta}}.
\end{align*}

\textbf{Last step} To finish the proof of the theorem, we need to consider the two cases, where either $\max_t p_t\le 0$ (NPVT) or $\max_t p_t> 0$ (not NPVT). We focus here on the NPVT because the other case is discussed in the proof of Theorem \ref{e:theo:delay}. 
If NPVT holds, we get for all $\tau$
\[
   \widehat{D}(\tau)\le   \max_{0 \le \ell < \tau} \bigg[\frac{1}{(\ell+1)^\beta \,\cdot\,T^{1/2-\beta}}\sum_{t=\tau-\ell}^\tau \frac{\hat p_t-p_t}{\hat \sigma_t}\bigg].
\]
For the right side (which is $\widetilde{D}(\tau)$), we have shown above weak convergence to $D$ (defined in \eqref{e:def:D}). Consequently
\begin{align*}
   &  \mathbb{P}\big(O = \textnormal{"DP violation detected"}\big) \\
    = &  \mathbb{P}\big(\max_\tau \widehat{D}(\tau)>q(\alpha) \big) \le \mathbb{P}\big(\max_\tau \widetilde{D}(\tau)>q(\alpha) \big)\\
    \to & \mathbb{P}\big(\max_\tau D(\tau)>q(\alpha) \big) = \alpha.
\end{align*}
This completes the proof.

\section{Proof of Corollary \ref{e:cor:stab}}

The corollary follows by an investigation of the proof of Theorem \ref{theo:main}. More precisely, since $c_n \to 0$, it follows that for all sufficiently large $n$ that $c_n <\sigma_1/2$. This means that in the definition of $Z_t$ it does not matter whether we have $\hat \sigma_t$ or the stabilized variance estimator $\tilde \sigma_t(c_n)$. The rest of the proof then proceeds without any changes.

\section{Proof of Theorem \ref{e:theo:delay}}

Without loss of generality we assume that the privacy violation we are searching happens already during the first observation period and thus $p_1>0$. Setting $\ell=\tau-1$ we get at time $\tau<\lfloor T\theta_1\rfloor$ that
\begin{align*}
    \widehat{D}(\tau) \ge &\bigg[\frac{1}{(\tau+1)^\beta \,\cdot\,T^{1/2-\beta}}\sum_{t=1}^\tau \frac{\hat p_t-p_t}{\hat \sigma_t}\bigg] \\
    + &\bigg[\frac{1}{(\tau+1)^\beta \,\cdot\,T^{1/2-\beta}}\sum_{t=1}^\tau \frac{p_t}{\hat \sigma_t}\bigg]\\
    =:&R_1+R_2.
\end{align*}
where $  \widehat{D}(\tau)$ is defined in \eqref{e:DT}. Obviously $R_1 \le \widetilde D(\tau)$, 
and we know that (uniformly in $\tau$) we have
\[
R_1 \le \max_\tau \widetilde{D}(\tau)=\mathcal{O}_P(1) 
\]
due to Lemma \ref{lem:_help}.
Thus, we study $R_2$. We know that $\sigma_1>0$ (otherwise $p_1>0$ is impossible). From \eqref{e:var} we have with probability going to $1$ the inequality that
\[
R_2 \ge \bigg[\frac{1}{(\tau+1)^\beta \,\cdot\,T^{1/2-\beta}}\sum_{t=1}^\tau \frac{ p_t}{\sigma_1/2}\bigg] \ge c_0\frac{\tau^{1-\beta} \sqrt{n}}{T^{1/2-\beta}},
\]
where $c_0$ is a sufficiently small, positive constant. The right side can be made arbitrary large if, for some sufficiently large constant $c_\tau$, we choose 
\[
\tau := c_\tau\bigg\lfloor \frac{T^{(1/2-\beta)/(1-\beta)}}{n^{1/(2(1-\beta))}}\bigg\rfloor
\]
and in particular delay times are of smaller (or equal) rate as the right side. To understand this conclusion, notice that $R_1$ is stochastically bounded, but we can make $R_2$ arbitrarily large and thus, we can make $R_1+R_2$ arbitrarily large, in particular larger than $q(\alpha)$, which then leads to detection.

\end{document}